\documentclass[usegraphicx,useAMS]{mn2e} 

\title[AGN-selected clusters as revealed by weak lensing]{AGN-selected clusters as revealed by weak lensing}

\author[Wold et al.\ ]
{Margrethe Wold$^{1,2}$, Mark Lacy$^{2}$, H{\aa}kon Dahle$^{3}$, Per B. Lilje$^{4}$, Susan E. Ridgway$^{5}$ \\
  \\
$^1$Stockholm Observatory, SCFAB, Roslagstullsbacken 21, SE-106 91 Stockholm, Sweden \\
$^2$SIRTF Science Center, Caltech, MS 220-6, 1200 California Bl., 
Pasadena, CA~91125, U.S.A. \\ 
$^3$NORDITA, Blegdamsvej 17, DK-2100 Copenhagen, Denmark \\
$^4$Institute of Theoretical Astrophysics, University of Oslo, P.O. Box 1029 Blindern, N-0315 Oslo, Norway \\
$^5$Department of Physics \& Astronomy, Johns Hopkins University, 3400 North 
Charles Street, Baltimore, MD 21218-2686, U.S.A.}

\date{ } 
\pubyear{2002}  

\begin{document}

\maketitle

\begin{abstract}
\noindent
As a pilot study to investigate the distribution of matter
in the fields of powerful AGN, 
we present weak lensing observations of deep 
fields centered on the radio-quiet quasar
E1821$+$643, the radio galaxy 3C 295, and the two radio-loud quasars, 
3C 334 and 3C 254. 
The host clusters of E1821$+$643 and 3C 295 are comfortably detected via 
their weak lensing signal, and we report the detection of a cluster-sized 
mass concentration in the field of the $z=0.734$ quasar 3C 254.
The data for the 3C 334 field are so far inconclusive. 
We find that the clusters are massive and have smooth mass distributions,
although one shows some evidence of past merger activity. The
mass-to-light ratios are found to be moderately high.
We discuss the results in light of the cooling flow
and the merger/interaction scenarios for triggering and fuelling AGN 
in clusters, but find that the data do not point unambiguously to neither 
of the two. Instead, we speculate that sub-cluster mergers may be responsible for 
generating close encounters or mergers of galaxies with the central
cluster member, which could trigger the AGN.
\end{abstract}


\section{Introduction}
\label{section:s1}

The causes of the dramatic cosmic evolution 
of AGN (e.g.\ Dunlop \& Peacock 1990;
Hartwick \& Schade 1990)
remain unclear, we have little knowledge of how and why 
AGN form, how quasar activity is triggered and sustained in a galaxy, and 
what extinguishes the activity. The findings that AGN are often 
associated with other galaxies is believed to indicate that
interactions between galaxies may play a role, since interactions 
are more likely to occur in a dense environment.  
Host galaxies and their close companions have been found to 
display disturbed morphology
(e.g.\ Bahcall et al.\ 1997; Canalizo \& Stockton 2001), 
also taken as a sign that interactions have taken place.
Even in cases where companion galaxies
appear undisturbed in the optical, radio images taken 
at redshifted 21cm (of nearby AGN) show long bridges or tidal 
tails of HI gas connecting the 
AGN with a close neighbour \cite{lh99}.

These observations have raised the question of the existence 
and the nature of a
physical link between the AGN and its environment. 
Two very different scenarios for this link have been suggested,
the merger/interaction and the cooling flow scenario. 

Ellingson, Green \& Yee \shortcite{egy91} found evidence for 
the merger/interaction
scenario when they made multi-object spectroscopy of galaxies associated
with optically luminous quasars. 
They found that
the richest quasar host clusters have abnormally low velocity dispersions
compared to those of clusters in general with similar optical 
richness. This is consistent with the idea that 
galaxy-galaxy interactions are related to the quasar phenomenon,
(e.g.\ Hutchings, Crampton \& Campbell 1984), 
since in dynamically young
clusters where the galaxies have low
velocities, the encounters between galaxies will be disruptive and
perhaps trigger or fuel quasar activity in one of the interacting
galaxies.   

Further support for this was found by Yee \& Green \shortcite{yg87}
and Ellingson et al.\ \shortcite{eyg91} in their studies of quasar
environments. Optically luminous quasars were found more frequently in
rich environments at  
$z\ga 0.6$ than at lower redshifts, suggesting that by $z\sim0.3$--0.4,
the quasars in rich clusters have faded. 
Since the time elapsed between
$z\sim0.4$ and $z\sim0.6$ corresponds to the dynamical time
scales of cluster cores, they proposed that dynamical
changes in the cluster environments were responsible for the fading
of quasars, perhaps as a result of lack of gaseous fuel after the
clusters virialize. 
Accordingly, highly luminous AGN should therefore not exist in 
rich clusters at low redshifts, but there are several examples 
which contradict this, demonstrating that the requirement of
non-virialized clusters is not sufficient.
One example is Cygnus A, a powerful FRII radio galaxy in a rich cluster
at $z=0.056$. Two other examples, which we study in this paper, are 3C 295 
and E1821$+$643. 3C 295 is a powerful FRII radio galaxy in a cluster 
at $z=0.46$, and E1821$+$643 is an optically luminous quasar at 
$z=0.297$ situated in a very rich cluster.

The cooling flow scenario has been discussed by e.g.\ 
Fabian et al.\ \shortcite{fabian86}, Fabian \& Crawford \shortcite{fc90}
and Bremer, Fabian \& Crawford \shortcite{bfc97}.
In this picture, cooling flows in sub-clusters at earlier epochs 
provide fuel for the AGN, but as sub-clusters merge toward the 
present epoch, the cooling flows are disrupted and the quasar 
activity switched off. Alternatively, the quasar may switch off
when the cooling flow gas is exhausted \cite{fc90}.  
Several observations have been made of extended emission-line
gas around powerful quasars which seem to support the cooling flow
model (e.g.\ Crawford \& Fabian 1989; Forbes et al.\ 1990;
Bremer et al.\ 1992).

However, it is not established yet whether extended emission line gas
indicates the presence of a cooling flow. For instance, the emission lines
could be produced by radio source shocks instead of gas cooling out of
the flow (e.g.\ Tadhunter et al.\ 2000). The standard cooling flow model
(Fabian 1994 and references therein) has received some criticism lately since
{\em XMM} observations of cooling flow clusters do not show
emission lines at temperatures below 1 keV as predicted (Peterson
et al.\ 2001; Tamura et al.\ 2001; Kaastra et al.\ 2001). This suggests that
the cooling flow model may need to be re-evaluated, and in many cases
mass deposition rates may turn out to be lower than originally thought.
Heating by AGN has been proposed as a possible cause of the apparent
lack of cold gas, but new {\em Chandra} data seem to indicate that the
coolest gas is close to the radio lobes, contrary to what is expected if
there is heating by the AGN (Fabian 2001).

If cooling flows were the chief mechanism for providing fuel to the most
powerful AGN in clusters, we should observe more rich clusters with central,
powerful AGN than we do. In fact, most rich AGN host clusters do not 
contain anything more powerful than an FRI source (like M87 in the Virgo 
cluster). In a sample of 260 rich-cluster radio galaxies studied by
Ledlow, Owen \& Eilek \shortcite{loe02}, there are almost no
FRII radio galaxies with luminosities at 1.4 GHz greater than
$10^{25.5}$ W\,Hz$^{-1}$.

Hall et al.\ \shortcite{hegy95} and Hall, Ellingson \& Green 
\shortcite{heg97} used X-ray data from {\em ROSAT} 
and {\em Einstein} to derive physical parameters, such as gas density
and temperature, of AGN host clusters.
Their aim was to distinguish between three different models for
the evolution of AGN in clusters, the young cluster scenario,
the cooling flow model, and a model of 
decreasing ICM density with increasing redshift (Stocke \& Perrenod 1981). 
Three of the five clusters in their sample were not detected, but
upper limits on their X-ray luminosity could be estimated. They found
that neither a cooling flow nor a decrease in ICM density could explain
all their observations, and proposed that strong interactions between
the AGN host and another galaxy in the cluster could be the 
only necessary mechanism to produce powerful AGN in clusters. 

In this paper, we investigate the link between four powerful AGN and their
environments by using weak lensing methods. The weak lensing technique allows us to 
probe the distribution of {\em total} mass (dark + luminous) in the AGN fields, 
and the mass distributions can be used to test the merger/interaction and the 
cooling flow scenarios.
The two scenarios make very different predictions about the
dynamical state of AGN host clusters, and by investigating the cluster
dynamics, it may be possible to find which, if any, of these
two mechanisms are responsible for the AGN--environment link. 

A cooling flow is expected to be associated with a single deep potential well in an 
evolved, virialized cluster \cite{esf92}, and in this case we expect a strong 
weak lensing signal and a mass distribution characterized by only a single peak. 
In contrast, 
if the AGN activity is related to galaxy interactions, these will be much more disruptive
if the encounter velocities are relatively low and comparable to the internal
velocity dispersions of the galaxies involved. Consequently, a cluster that
is still forming will be preferred, as in such a cluster there is an increased
probability for low-velocity encounters. A dynamically young cluster is
expected to have an irregular mass distribution, possibly with several sub-clumps
\cite{schindler01}. 

Our method of cluster selection is based on the detection of weak gravitational
shear and the presence of an AGN. This is different from the usual methods of detecting
clusters, by optical richness or X-ray luminosity.
Whereas optical and X-ray cluster surveys 
are biased toward the most baryon-rich systems, clusters selected by their 
weak shear signal are biased toward the most massive systems in terms of dark+luminous
mass. If there is a large range in the baryonic-to-dark matter ratio in clusters, 
clusters with high mass-to-light (hereafter M/L) ratios will have been missed in optical and
X-ray surveys. Although we present a case-by-case study here, it is nevertheless 
interesting to compare the masses and the M/L ratios of the AGN host clusters to those
of clusters detected on the basis of optical richness or X-ray luminosity.

The outline of the paper is as follows. In Section~\ref{section:s2} we 
describe the 
selection of the targets and review some of their properties. 
Section~\ref{section:s3} describes the observations and 
the data reduction. 
Before the weak lensing analysis is performed,
we investigate the colour-magnitude diagrams of the clusters in 
Section~\ref{section:s4}. The purpose of this is to locate the colour-magnitude
relation in the known clusters, E1821$+$643 and 3C 295, and thereby identify 
cluster galaxies that do not contribute to the weak lensing signal. We also
investigate the colour-magnitude diagram of 3C 254 and discuss whether there
is evidence for a red sequence of early-type cluster galaxies in this field.
A brief introduction to the weak lensing method and how we measure the
gravitational shear is given in 
Section~\ref{section:s5}. We follow the standard approach for ground-based
data and use the
Kaiser, Squires \& Broadhurst \shortcite{ksb95} formalism to correct for 
PSF anisotropies using the {\sc imcat} software, and the approach by 
Luppino \& Kaiser \shortcite{lk97} to calibrate galaxy ellipticities to 
gravitational shear.
In Section~\ref{section:s6}, we present the results of the analysis by 
showing maps of the projected surface mass density in the fields, 
plots of
the shear profiles in the clusters, and estimates of masses and M/L ratios.
The results are discussed in Section~\ref{section:s7}, and conclusions are 
drawn in Section~\ref{section:s8}.

Our assumed cosmology has $h=H_{0}/100$ km\,s$^{-1}$\,Mpc$^{-1}$,
$\Omega_{0}=1$ and 
$\Omega_{\Lambda}=0$ 
\footnote{For this geometry, one arcsec is equivalent to
2.7$h^{-1}$ kpc at $z=0.297$, 3.4$h^{-1}$ kpc at $z=0.46$, 3.7$h^{-1}$ kpc at 
$z=0.555$ and 4.0$h^{-1}$ kpc at $z=0.734$, the redshifts of the targets in this study.}.


\section{The sample} 
\label{section:s2}  

We selected four powerful AGN with extended X-ray emission for this study. 
Two of the targets are well-known cases of an AGN associated
with a galaxy cluster, the radio galaxy 3C 295 and the radio-quiet quasar E1821$+$643.
The other two targets are the radio-loud quasars 3C 334 and 3C 254 which were
found to have significant extended X-ray emission in moderately deep {\em ROSAT}
pointings (Hardcastle \& Worrall 1999; Crawford et al.\ 1999). 
For the former two sources, we know
that the extended X-ray emission originates in the hot intra-cluster 
medium of the host clusters, but in the latter two cases no 
host clusters are known. 
Their extended X-ray emission is however an
indicator of clustered galaxy environments provided that the point-like X-ray 
emission from the AGN has been properly subtracted, and that the 
extended component does not originate in processes associated with 
the AGN itself. 

\subsection{3C 295}

This is a powerful FRII radio galaxy at $z=0.46$ with compact radio structure.
In the optical, the radio galaxy is classified as a cD,
and its host cluster has been extensively studied both 
at optical (Mathieu \& Spinrad 1981; Dressler et al.\ 1997; 
Smail et al.\ 1997a,b; Standford, Eisenhardt \& Dickinson 1998;
Thimm et al.\ 1998) and X-ray wavelengths (Henry \& Henriksen 1986; 
Mushotzky \& Scharf 1997; Neumann 1999; Allen et al.\ 2001). 
It is not a particularly optically-rich
cluster, most likely of Abell class 1
(Mathieu \& Spinrad 1981; Hill \& Lilly 1991), but its velocity dispersion is 
high, $\sim 1630$ km\,s$^{-1}$ \cite{dressler99}. 

In X-rays, it appears to be a dynamically evolved and relaxed cluster with a 
high bolometric luminosity of $L_{\rm X}\approx6.5\times10^{44}h^{-2}$
\cite{neumann99}.
It was suggested 
by Henry \& Henriksen \shortcite{hh86} that the cluster hosts a cooling 
flow, later confirmed by Neumann \shortcite{neumann99}. 
Allen et al. \shortcite{allen01} have studied the cluster with the
{\em Chandra} X-ray satellite and find that the 
cooling flow is 2--4$h$ Gyr old, with a mass 
deposition rate of 70$h^{-2}$ M$_{\odot}$\,yr$^{-1}$.

3C 295 was one of the first two clusters in which a weak gravitational 
lensing signal
was detected \cite{tvw90}, and Smail et al.\ \shortcite{smail97a} have 
performed weak lensing analysis on this field using HST WFPC2 data. 
The cluster displays signs of strong lensing and a quadruple lens has also
been found (Fisher, Schade \& Barrientos 1998; Lubin et al.\ 2000). 
Since this cluster is well-studied, it serves as a 
consistency check on our techniques. 

\begin{table*} 
\begin{minipage}{11.3cm}
\caption{Data obtained during the three observing runs  
from 1998 to 2000. We list the seeing and the completeness limit in the combined
images, but for 3C 295 and E1821$+$643, only using the combined $I$-band
images from 1999, since these were used for the weak
lensing analysis.
The completeness limits are based on the turn-over in the number counts 
(from object detection with a 4$\sigma$ detection threshold).} 
\begin{tabular}{llllllll} 

Field & $z$ &  Filter & \multicolumn{3}{l}{Exposure time (s)} &  Seeing   & Completeness\\
      &     &         & 1998 & 1999 & 2000                &  (arcsec) & limit \\ 
      &     &         & &  &               &  &  \\ 

E1821$+$643  & 0.297 & $I$ & 7500 & 10800& --  & 0.83 & 24.5 \\ 
            &       & $V$ & 1800  & 7200  & --  & 1.00 & 25.5 \\       
            &       & $B$ & --    & 15600 & --  & 0.85 & 26.5 \\  

3C 295       & 0.460 & $I$ & 9300  & 10800 & --  & 0.76 & 25.0 \\ 
            &       & $V$ & --    & 10800 & --  & 0.95 & 25.5 \\

3C 334       & 0.555 & $I$ & --    & -- &  13790 & 0.62 & 23.5 \\ 

3C 254       & 0.734 & $I$  & --    & -- &  21747 & 0.50 & 24.0\\
            &       & $V$ & --    & -- &  15500 & 1.10 & 26.5\\

\end{tabular} 
\label{table:tab1} 
\end{minipage}
\end{table*}  

\subsection{E1821$+$643}

This is an optically luminous quasar at $z=0.297$ residing in 
a giant elliptical host galaxy (Hutchings \& Neff 1991;
McLeod \& McLeod 2001) at the centre of a very rich Abell class 
$\ga 2$ cluster (Schneider et al.\ 1992; Lacy, Rawlings \& Hill 1992). 
The quasar is detected at radio wavelengths, has a compact core, a 
one-sided jet,
and a radio luminosity of $L_{\rm 5 GHz}\approx 1.2\times10^{23}h^{-2}$ 
W\,Hz$^{-1}$\,sr$^{-1}$ (Lacy et al.\ 1992; Blundell et al.\ 1996), but 
is classified as radio-quiet because of its high optical 
luminosity, $M_{B}\approx-27$.

The surrounding cluster was first noted in the optical by Hutchings \& Neff \shortcite{hn91}
who believed it was background to the quasar, but Schneider et al.\ \shortcite{schneider92}
took spectra of eight galaxies in the field, and found that six of 
them belonged to 
a cluster at the quasar redshift.
Subsequently, spectra of several galaxies in this field
have been obtained in studies of Ly$\alpha$ absorption systems
(Le Brun et al.\ 1996; Tripp et al.\ 1998), and using 
published spectroscopic redshifts of 42 galaxies the cluster velocity dispersion 
is estimated to 1182$\pm$19 km\,s$^{-1}$ (kindly provided to us by B. Holden, 
using the biweight algorithm of
Beers, Flynn \& Gebhardt \shortcite{bfg90} and a jackknife technique
to estimate errors).

Even though it is one of the most X-ray luminous clusters known, with
a bolometric luminosity of $L_{\rm X} \sim 10^{45}h^{-2}$ 
erg\,s$^{-1}$ \cite{saxton97}, it went undetected in X-rays for 
a long time due to the
presence of the dominant quasar point source. But recently, 
Saxton et al.\ \shortcite{saxton97} and 
Hall et al.\ \shortcite{heg97} reported detections of excess 
X-ray emission around the quasar in {\em ROSAT} data.
The X-ray data gave no firm evidence for a cooling flow,
and observations made by Fried \shortcite{fried98} 
of extended emission-line gas around the quasar
suggest that no cooling flow is present since 
the [{\sc oiii}]/[{\sc oii}] emission-line ratio does not  
increase outward as expected in a cooling flow.

It has been pointed out that this quasar has many of the
properties typical of radio-loud quasars. It has a giant elliptical host, 
lies in a rich cluster, and the 
[{\sc oiii}] luminosity in the extended emission-line region 
is more typical of that found in radio-loud quasars (Fried 1998, and 
references therein). 
However, E1821$+$643 fits well in with quasars from the 
FIRST Bright Quasar Survey (FBQS, Gregg et al.\ 1996; White et al.\ 2000) 
in the radio luminosity--optical luminosity plane. Lacy et al.\ 
\shortcite{lacy01} used the FBQS to show that there is a continuous variation
of radio luminosity with black hole mass for quasars. Using the
H$\beta$ line width of E1821$+$643 \cite{kolman93}, an estimate 
of the black hole mass can be made (Kaspi et al.\ 1999), and we find that the
black hole in E1821$+$643 has a mass of $\sim 1\times10^{9}h^{-1}$ 
M$_{\odot}$ (for details, see Lacy et al.\ 2001). 
With this black hole mass and its radio luminosity, it fits very well
the radio-luminosity -- black hole mass relation of Lacy et al.\ and 
is therefore similar to the FBQS objects. E1821$+$643
might not be such a special case after all, and is probably 
typical of highly luminous radio-quiet quasars. 

\subsection{3C 334 and 3C 254}

These are two powerful steep-spectrum radio-loud quasars at $z=0.555$ and 
$z=0.734$, respectively, and are the most distant quasars in 
our study. They were included in the sample because there exists evidence that 
they are surrounded by extended X-ray emission.
At rest frame 0.1--2.0 keV, Crawford et al.\ \shortcite{crawford99}
find X-ray luminosities of 
0.8--2.0$\times$10$^{44}h^{-2}$ and 1.3--2.3$\times$10$^{44}h^{-2}$ erg\,s$^{-1}$ for the 
emission around 3C 334 and 3C 254, respectively. 
If the extended emission is interpreted as thermal 
emission from a hot intra-cluster medium, the quasars may lie in 
rich clusters since the typical luminosity of 
Abell class 1 clusters is a few times $10^{44}h^{-2}$ erg\,s$^{-1}$
\cite{bh93}.

There is further circumstantial evidence that 3C 334 and 3C 254 are 
associated with galaxy clusters. Optical images reveal an apparent overdensity
of galaxies around 3C 254 \cite{bremer97}, and Hintzen \shortcite{hintzen84}
find a number of nearby companion galaxies to 3C 334, although 
Yee \& Green \shortcite{yg87} do not find any significant clustering in this field.
Further suggestions that 3C 334 and 3C 254 lie in clusters 
come from observations of extended oxygen line-emission
around the quasars where the pressures inferred from the oxygen lines are 
comparable to pressures in the intra-cluster medium of nearby clusters
(Crawford \& Fabian 1989; Forbes et al.\ 1990; Crawford \& Vanderriest 1997).
It is argued by Forbes et al.\ and Crawford \& Vanderriest that 3C 254 might
be embedded in a very massive cooling flow with a mass deposition rate of 
up to $250h^{-2}$ M$_{\odot}$\,yr$^{-1}$ within 20$h^{-1}$ kpc of the quasar. 
If the quasar is indeed associated with a massive cooling flow, one might expect
that the cluster around 3C 254 be very massive as well. 


\section{Observations and data reduction}
\label{section:s3}

Imaging observations were made with the 2.56-m Nordic Optical Telescope (NOT) 
using the ALFOSC 
(Andalucia Faint Object Spectrograph and Camera) instrument equipped with a 2k$\times$2k 
Loral CCD with pixel scale 0.189 arcsec, resulting
in a 6.5$\times$6.5 arcmin$^2$ FOV.
The NOT/ALFOSC is a well-suited instrument for weak lensing studies of clusters, both because
the field of view is relatively large, and because the NOT site has excellent seeing conditions.
A wide field of view ensures a large number of faint background 
galaxies to measure ellipticities on, thereby improving the statistics, and good seeing
conditions minimize the noise in the determination of galaxy ellipticities.

Since the seeing is improved at longer wavelengths, we 
used the $I$-band images for measuring galaxy shapes. We therefore dedicated
the time when the FWHM seeing was $\la 0.8$ arcsec to $I$-band imaging, and
observed in $V$-band when the seeing increased to 0.9 arcsec.
The E1821$+$643
cluster was also observed in the $B$-band since there is a planetary
nebula in this field (K1-16, 88 arcsec NW of the quasar) which 
is relatively bright in $V$, but less bright in $B$, since most of the
emission comes from {\sc [oiii]}$\lambda \lambda$4959,5007.

The integrations were divided into exposures of 600 s each in the $B$- and $V$-band, and 
300 or 450 s in the $I$-band, offsetting the telescope 10--15 arcsec between each exposure 
to avoid bad pixels falling consistently on the same spot.

We had in total three observing runs to obtain the data for this project, as 
summarized in Table~\ref{table:tab1}. On the 1998 run,  
there were clouds and cirrus on several of the nights, and
the conditions were therefore generally poor for doing 
weak lensing observations.
A backup programme was done instead, but occasionally during 
clear periods, we could return to the weak lensing programme.
During the 1999 and 2000 runs, the weather was clear and the conditions
photometric.

The data reduction was performed using standard {\sc iraf} routines for bias subtraction and 
flatfielding. 
For the flatfielding, we used twilight flats, but we also flatfielded the 
$I$-band data with a skyflat formed by taking the median of nine object frames. 
By dividing with the skyflat we were able to remove a fringing pattern 
before the images were combined. The fringing pattern 
is additive and should ideally be subtracted from the images, but we found that division with
the skyflat removed the fringes much better. For 10 per cent fringes, the 
error in the photometry will be 
$\pm0.1$ mag, but since our analysis is more critical to measuring galaxy shapes accurately, 
we chose a flat background level at the expense of accurate photometry.

After bias subtraction, flatfielding and fringe removal, the images 
were aligned and coadded using
the {\sc iraf} tasks {\em imalign} and {\em combine} with combine 
operation set to average.  
In order not to confuse the detection algorithm in the {\sc imcat} software,
we also flattened out large-scale variations in the sky level
by subtracting off a model of low-frequency variations 
in the background level. 

The photometric zero points were determined by taking images of 
standard star fields \cite{landolt92}, and the atmospheric extinction 
was estimated by observing the standard fields at different airmasses.
Galactic reddening was corrected for by looking up the amount of extinction 
in the NED\footnote{Nasa Extragalactic Database}. The NED values are based 
on reddening maps by 
Schlegel, Finkbeiner \& Davis \shortcite{sfd98}, and the extinctions are calculated
assuming an $R_{\rm V}=3.1$ extinction curve.  

When we aligned the images, we discovered that the frames from 1999 were rotated with 
respect to those taken in 1998 by
0.242$\pm$0.005 degrees. The small rotation caused
problems when we combined the data from 1998 and 1999, as periodically 
varying ellipticities
were introduced when one set of images was 
rotated with respect to the other. We therefore performed 
the weak lensing analysis on just the 1999 $I$-band data, since these were both
deeper and taken during better conditions than those from 1998. 
The full data set was instead
used when we estimated the light from the galaxies in order
to determine the M/L ratios of the clusters.

It is seen in Table~\ref{table:tab1} that the
combined $I$-band images have a better seeing ($\la 0.8$ arcsec) than those 
taken in the $V$-filter ($\approx 1$ arcsec), and with the exception of 3C 254,  
we used the $I$-band images for the weak lensing analysis. 
Since the $V$ image of 3C 254 is very deep compared to 
the $I$ image, it was more feasible to use this for 
the weak lensing despite it being the poorer-seeing image.
The $I$-band image of 3C 334 is seen to only 
reach $I\approx23.5$, and more
data is needed on this field to be able to perform a 
reliable weak lensing analysis.

We used two software packages for object detection.
For the weak lensing analysis, we utilized the {\sc imcat} software \cite{ksb95}
which is optimized for shape measurements on faint galaxies 
and weak lensing.
During other steps of the analysis when more accurate galaxy photometry was 
wanted, we used SExtractor \cite{ba96}, which is specifically designed for
photometry.


\section{Identifying the early-type cluster population}
\label{section:s4}

\begin{figure}
\includegraphics{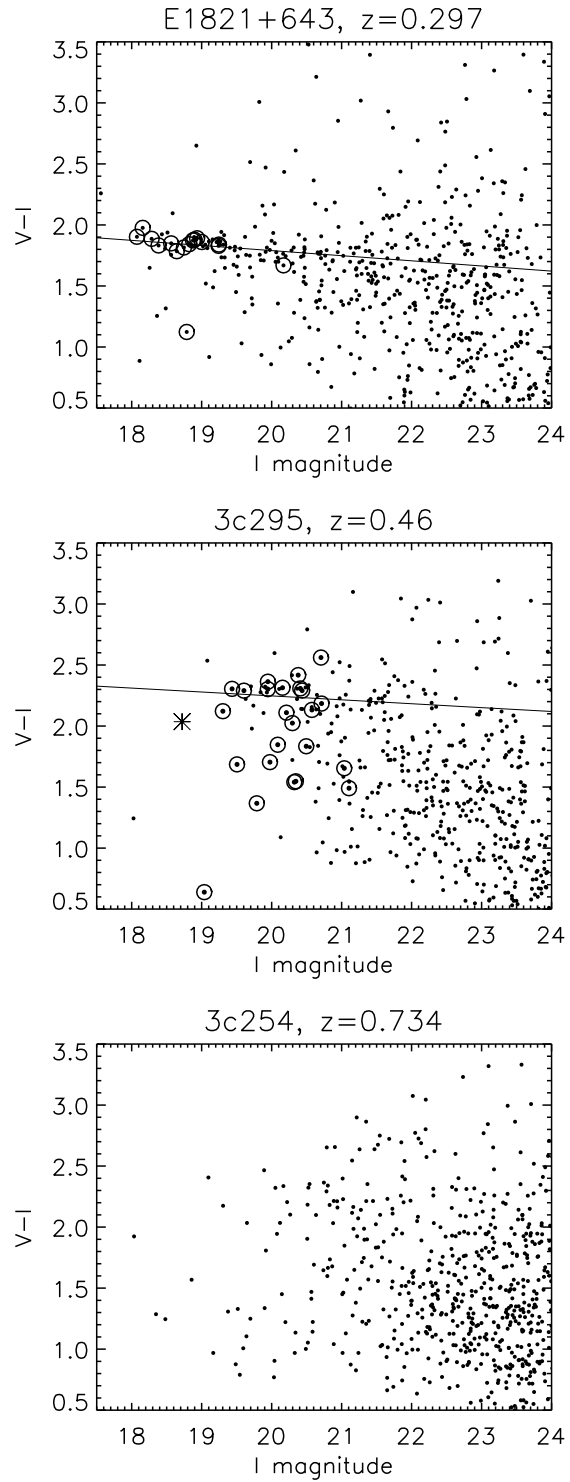}
\caption{Colour-magnitude diagrams of galaxies in the fields of 
E1821$+$643, 3C 295 and 3C 254. 
The encircled points correspond to spectroscopically 
confirmed cluster members by Dressler et al. \protect\shortcite{dressler99} (for 3C 295),
Schneider et al. \protect\shortcite{schneider92}, Le Brun et al.\ \protect\shortcite{lbb96}
and Tripp et al.\ \protect\shortcite{tls98} (for E1821$+$643). The radio galaxy 3C 295
is marked with an asterisk.}
\label{figure:fig1}
\end{figure}

In this section we investigate the colours of the galaxies in order
to identify galaxies that are likely to belong to the clusters. 
We wish to identify probable cluster members for two reasons. First, because it allows us to exclude them
from the weak lensing analysis, thereby improving the signal-to-noise.
Only galaxies behind the clusters can become gravitationally distorted, so 
both cluster members and foreground galaxies will contaminate the weak 
shear signal. Second, the identified cluster members can be used to 
estimate the optical light from the cluster, which will be useful
later for constraining the M/L ratios.

To do this, we studied the colour-magnitude diagram of
the quasar fields in order to locate the locus of possible cluster galaxies.
Early-type galaxies in clusters are known to form a sequence in the colour-magnitude
plane, often referred to as the colour-magnitude relation, or the red sequence.  
The sequence is remarkably tight with little scatter and is seen in both
nearby and more distant clusters (e.g.\ Stanford et al.\ \shortcite{sed98}).
Ideally, we would like to know the redshifts of all the galaxies in the
fields, but this would require deep imaging in two--three more filters so that 
photometric redshifts could be obtained.
Using the colour-magnitude relation is, however, 
a crude photometric redshift method,
since it is based on the strong 4000 {\AA} breaks in the spectral energy distributions
of early-type galaxies.

To evaluate colours of the galaxies we used magnitudes determined within a 
2.6 arcsec aperture, corresponding approximately to 3 times the seeing. 
At the redshifts of the clusters studied here, the aperture corresponds
to $\approx$ 7--10$h^{-1}$ kpc.  
The colour-magnitude diagrams for the three fields that were observed
in $V$ and $I$ are shown in Fig.~\ref{figure:fig1}.
The colour-magnitude relation in the E1821$+$643 cluster is identified 
as the almost horizontal band stretching from $I\approx18$ to 21, 
where it starts to blend with the background population. In the 3C 295 cluster,
the relation is less well-defined, because the cluster is poorer and lies
at a higher redshift. In the case of 3C 254, we can use the colour-magnitude
diagram to search for a cluster by looking for the signature of 
a red sequence. We return to this in Section~\ref{section:s41}.

Since we are not able to exclude non-cluster galaxies by means of 
spectroscopic redshift, there will be much scatter around the red sequence
relation in each cluster. 
In order to determine the zero points and the slopes of the colour-magnitude
relations 
in the 3C 295 and the E1821$+$643 clusters, 
we therefore used a sigma clipping 
algorithm on regions in the colour-magnitude diagram where it was
clear that the red sequences lay. 
In the case of E1821$+$643, this region was defined at $18.0 < I < 20.0$
and $1.5 < V-I < 2.5$ for the $V-I$ relation, 
and at $20 < V < 22$ and $1.3 < B-V < 2.0$ for the $B-V$ relation.
For the 3C 295 cluster, the region was constrained to 
$19 < I < 21.5$ and $2.0 < V-I < 2.5$.
After the sigma clipping, we fitted a straight line to the 
remaining galaxies yielding
the slope and the zero point. The results are listed in 
Table~\ref{table:tab2} in the form 
${\it colour} = {\it zpt} + {\it slope}\left(m-m_{0}\right)$, where {\it zpt} 
is the colour at the apparent magnitude $m=m_{0}=21$.
The relations we derive agree with those found by e.g.\ Ellis et al.\ 
\shortcite{ellis97} and those predicted by Kodama et al.\ \shortcite{kodama98}. 

\begin{table}
\caption{The colour-magnitude relation in the clusters in terms of 
slope and zeropoint. The zero point is given as the colour at an
apparent magnitude of $m=21$. The error in the slope was calculated 
assuming errors in $V-I$ and $B-V$ of 0.1.}
\begin{tabular}{lllll}
Cluster     & {\it colour} & $m$   & {\it zpt} & {\it slope} \\
            &        &     &                         &     \\
E1821$+$643  & $V-I$ & $I$ & 1.75 & $-$0.04$\pm$0.02 \\
E1821$+$643  & $B-V$ & $V$ & 1.50 & $-$0.04$\pm$0.02 \\
3C 295       & $V-I$ & $I$ & 2.22 & $-$0.03$\pm$0.03 \\
\end{tabular}
\label{table:tab2}
\end{table}

In order to reject galaxies belonging to the clusters
in the weak lensing analysis, we made cuts in 
magnitude and colour based on the derived colour-magnitude relations.
First, bright galaxies at $I\leq21$ 
were excluded irrespective of colour to ensure that also 
bright foreground galaxies were removed (a magnitude of $I=21$
corresponds roughly to M$^{*}$ at $z\approx0.7$).
Thereafter, galaxies lying within 1$\sigma$  of 
the $V-I$ colour-magnitude relation at $I>21$ were rejected 
($\sigma$ is the dispersion in colour after the sigma clipping 
in the regions defined above, typically 0.1--0.3).

\begin{figure} 
\includegraphics{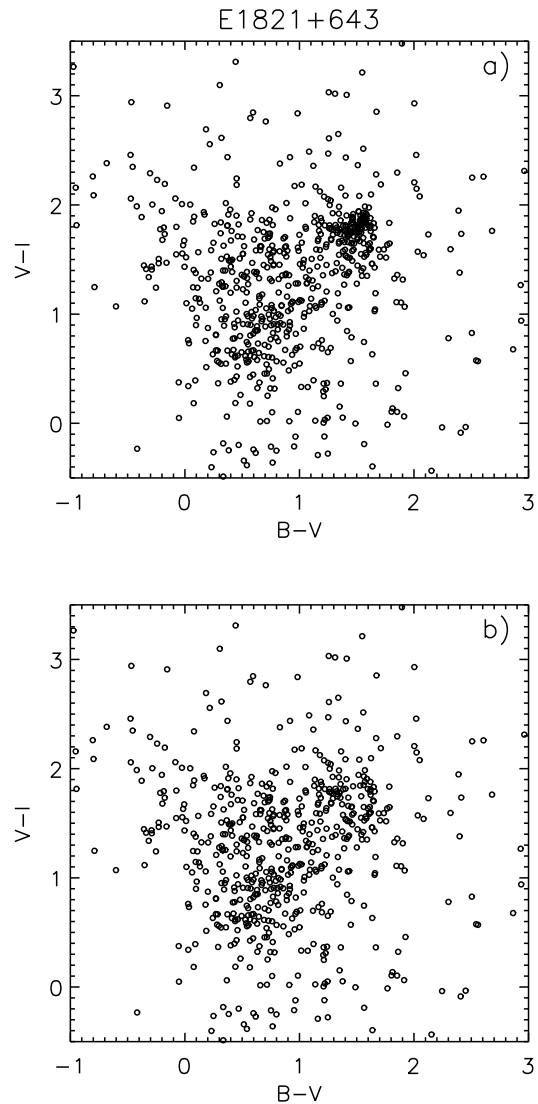} 
\caption{Colour-colour diagram of galaxies in the E1821$+$643 field. In panel a)
we have plotted all galaxies in the field, and panel b) shows the galaxies that are 
left for the weak lensing analysis after rejection of early-type cluster
galaxies.} 
\label{figure:fig2}
\end{figure}

For the E1821$+$643 cluster, we also used the $B-V$ information
since the early-type galaxies in this cluster are confined to a 
small region in the colour-colour plane.
This can be seen in the upper panel of Fig.~\ref{figure:fig2}, where the 
red sequence galaxies stand out as an overdensity at $B-V\approx1.5$
and $V-I\approx1.8$. The lower panel shows the galaxies that
are left after the rejection of early-type cluster galaxies.

For 3C 254, we rejected galaxies at $V\leq22$ in the weak 
lensing analysis, but did not
perform any colour-cuts since no cluster is known {\it a priori}
in this field. Since only $I$-band data were taken of the 
3C 334 field, we excluded galaxies at $I<20$.


\subsection{The distribution of red galaxies in the 3C 254 field}
\label{section:s41}

There are some features in the 3C 254 colour-magnitude
diagram that caught our eye and made us speculate 
(prior to the weak lensing analysis) whether this field contains two 
clusters.
Ignoring the brighter galaxies at 
$V-I\approx1$, there seems to be an excess of 
red galaxies at $I\approx19$--21 with colours 
$V-I\approx2.3$ resembling a red sequence. 
There is also an apparent excess of galaxies at $I\approx20.5$--22.3
and $2.5 < V-I < 3.0$ which is not seen in the other two clusters.
Since this is the expected colour of early-type galaxies at $z \ga 0.7$,
it could be the red sequence in a cluster at the quasar redshift.

\begin{figure*}
\includegraphics[scale=0.8]{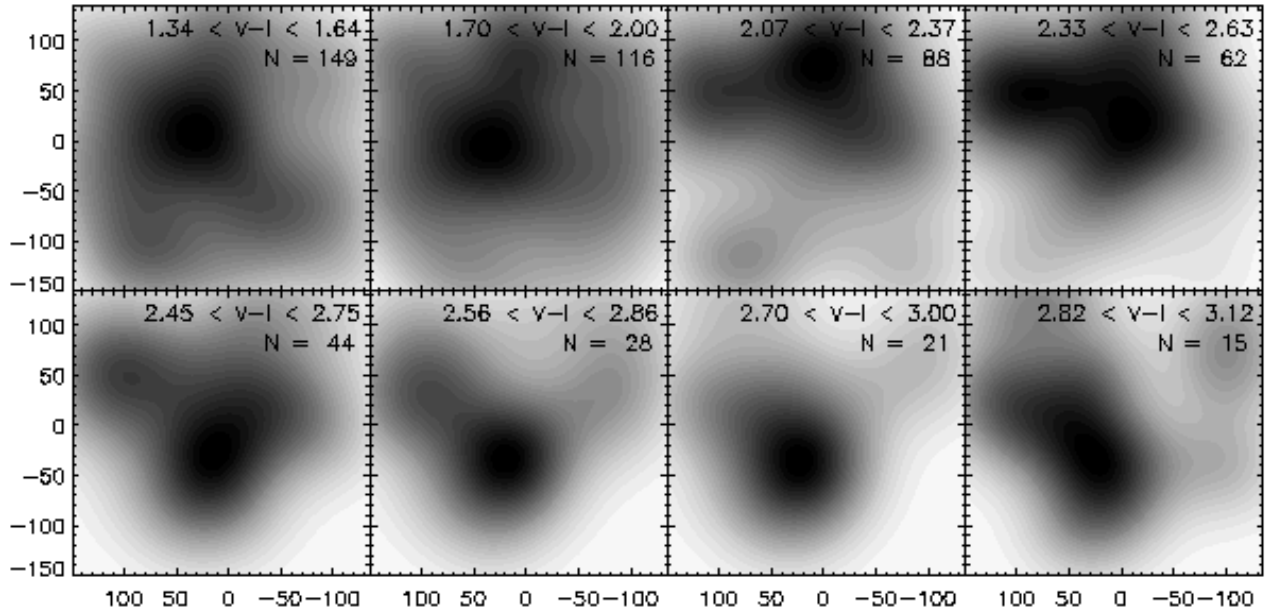}
\caption{The results of the cluster-finding algorithm in the 3C 254 field.
The panels show the smoothed galaxy surface density distribution as a function
of predicted colour-magnitude relations at different redshifts. The redshifts are,
from top-left to bottom-right, $z=0.2$ to 0.9 in steps of $\Delta z = 0.1$.
The numbers on each plot show the colour range we assumed that observed $L^{*}$
elliptical galaxies would have at each redshift, as well as the number of galaxies
in each colour range (see text for details).  
Axis units are offset from the quasar in arcsec, and there are 
30 greylevels in each panel, defined to span the range from maximum to minimum.
North is up and east is to the left.}
\label{figure:fig3}
\end{figure*}

In order to investigate this in more detail, we derived the 
surface density distribution of the galaxies as a 
function of colour.
We followed the approach by Gladders \& Yee \shortcite{gy00},
who have constructed an algorithm to search for clusters in two-filter
data (the `Red-Sequence Cluster Survey'). 
Galaxies on a cluster red sequence will be redder than 
any normal foreground galaxies provided the two filters straddle 
the 4000 {\AA} 
break in the rest frame of the cluster. In this manner, $V$- and $I$-filters
can be used to search for clusters at $z\approx0.4$--0.9.
We did not implement the whole algorithm since we are not doing a 
large blank field survey, but merely a simplified version of 
the first part which 
assigns probabilities to the galaxies according to whether they are 
likely
to lie in a certain colour interval. 
If there is a concentration in the surface density of galaxies 
in a particular colour interval, it classifies as a cluster candidate.
The colour interval will then be an indirect indicator of the
cluster redshift.

To implement this, we defined colour-magnitude relations between
$z=0.20$ and 0.95 for every $\Delta z = 0.05$.
The zero points of the predicted 
relations were taken to be equal to the $V-I$ colour of an $L^{*}$ elliptical galaxy
at the redshift in question.
This was kindly provided by T. Dahl{\'e}n from his photometric redshift code,
which convolves different galaxy templates with the NOT filters \cite{dfn02}.
The slope of the colour-magnitude relation was taken to vary slightly with
redshift, as predicted by Kodama et al.\ \shortcite{kodama98} for their elliptical 
evolution model with a formation redshift of $z_{f}=4.5$. 
We took the scatter
about the colour-magnitude relation to be $\Delta\left(V-I\right)=0.3$ so that it
roughly matches the intrinsic scatter and the photometric errors, the latter 
clearly dominating.

We identified galaxies likely to belong to the 
defined red sequences by selecting those that had a certain probability
of lying in the colour interval, $\left<c_{1},c_{2}\right>$, associated with each 
sequence. This probability was evaluated as the integral from $c_{1}$ to $c_{2}$ of a
normal distribution with mean equal to the measured $V-I$ and 
dispersion equal to the measurement error in $V-I$ (for details, see Gladders \& Yee 2000).
Thereafter, we evaluated the surface density distribution of galaxies above a
given probability and smoothed it with a Gaussian of width
200 pixels (the same as we use in the mass maps). The result is shown in 
Fig.~\ref{figure:fig3}, where each panel corresponds to a colour-magnitude relation 
at a given redshift. In order  
to find a potential cluster in these plots, we looked for well-defined peaks 
in the surface density within a narrow colour range.
 
We find two distinct peaks in this figure. In moving from bluer toward
redder colours, the first peak turns up at $2.07 < V-I < 2.37$,
70--80 arcsec N of the quasar. If this is a cluster red sequence, it is predicted
to have a redshift of $z\approx0.4$. The second isolated peak shows up in the 
two panels spanning the colour range $2.56 < V-I < 3.00$, and  
if these galaxies are early-type members of a cluster, the predicted redshift 
is $z\approx0.7$--0.8. 
It is thus possible that the second peak might be a cluster associated with the quasar.
Interestingly, the cluster does not seem to be centered on the
quasar, but rather 40--50 arcsec to the S.
If the quasar is associated with a strong cooling flow (Forbes et al.\ 1990; 
Crawford \& Vanderriest 1997) in a cluster, we would expect it to lie at the 
cluster centre where the gravitational well is deep. 
However, it cannot be ruled out by our data that the cluster might have a 
slightly higher redshift than 0.734, but it is 
unlikely that the redshift is higher than 0.85--0.9.


\section{Weak lensing analysis}
\label{section:s5}

A cluster of galaxies has a high enough mass density to cause deflection  
of light rays from distant galaxies behind it, thereby distorting their shapes.
Depending on the surface mass density
of the cluster and the alignment of the background galaxies with the cluster centre of 
mass and the observer, different levels of distortions are seen -- from 
giant arcs and arclets to only weakly altered galaxy shapes.

Giant arcs are produced in the strong lensing regime when there is good alignment
and the surface mass density in the cluster centre is equal to or larger than the critical 
surface mass density, $\Sigma_{\rm c}$,
\begin{equation} 
\Sigma_{\rm c}=\frac{c^{2}}{4 \pi G} \frac{D_{\rm s}}{D_{\rm d}D_{\rm ds}}. 
\label{equation:eq1}
\end{equation}
\noindent  
Here $D_{\rm s}$ and $D_{\rm d}$ are the angular diameter distances to the source
(background galaxy) and the deflector (the cluster), respectively, and 
the deflector--source angular diameter distance is $D_{\rm ds}$. 
The weak lensing regime is characterized by the condition $\kappa = \Sigma/\Sigma_{\rm c} \ll 1$, 
and in this case the background galaxies are only weakly distorted.
To detect this effect, we need to measure the ellipticities of a large number of galaxies
and look for a systematic shift, in particular for a tangential alignment
of the galaxy shapes around the cluster centre.  

As opposed to giant arcs, weak lensing occurs in every cluster, but is 
difficult to quantify  
because the background galaxies themselves have an intrinsic distribution of ellipticities. This is the main source of noise in weak lensing analysis.
In addition to this, the faint galaxies are smeared by the seeing 
PSF which will circularize the galaxy images
and suppress the weak shear signal. 
There may also be anisotropies in the PSF causing it to change as a 
function of position in the image. 
This might be caused by e.g.\ camera distortions, guiding errors, 
or strong winds giving rise to vibrations in the mirror or 
shaking of the telescope. Also, the strain in the instrument
caused by its own weight changes direction as the telescope
tracks an object across the sky, and this can have an
effect on the shapes of the galaxies.

Fortunately, well-developed techniques for measuring galaxy ellipticities
and correcting for PSF smearing and anisotropy exist.
Here, we use the
{\sc imcat} software by N. Kaiser in which the techniques described by 
Kaiser et al.\ \shortcite{ksb95} are implemented. 
{\sc imcat} detects objects by smoothing the images with a range
of Mexican hat filters with different radii. The detection
with the highest significance is entered into the catalogue along
with parameters such as magnitude, the radius $r_{\rm g}$ of the smoothing filter 
which maximized the detection significance, and ellipticity.
Galaxy ellipticities are evaluated as 
\begin{equation} 
e_{1}=\frac{Q_{11}-Q_{22}}{Q_{11}+Q_{22}} \hspace{1cm} e_{2}=\frac{2Q_{12}}{Q_{11}+Q_{22}}, 
\end{equation}
\noindent 
where the elements of the Q matrix are formed from the weighted second 
moments of the surface brightness.

In each image, we examined the result of the {\sc imcat} detection 
process, and made masks when necessary to eliminate spurious objects.
In order to include only galaxies likely to have well-determined
shapes in the weak lensing analysis, we accepted only 
galaxies that satisfied the 
following criteria: 1) $r_g$
larger or equal to the stellar radius, but less than
six times the stellar radius, 2) significance of detection ($\nu$) greater
than six, and 3) corrected ellipticity less than three (see Section~\ref{section:s51}
for a description of the ellipticity corrections). 
Stars with $I<22$ were separated from galaxies 
on the basis of a plot of half-light radius versus
magnitude,
as shown in Fig.~\ref{figure:fig4}, and removed from the catalogues.
No attempt was made to separate stars from galaxies at $I > 22$. 
At these faint magnitudes, the surface density of galaxies is 
larger than that of stars by a factor of $> 20$ (Reid et al.\ 1996), so 
the inclusion of faint stars will not affect our analysis.
Bright foreground galaxies and 
probable cluster members were also taken out of the catalogues
by making the cuts in magnitude and 
colour as described in Section~\ref{section:s4}.
The number of galaxies left for the weak lensing analysis
was typically $\sim$1000, see Table~\ref{table:tab3}.


\subsection{Shear measurements}
\label{section:s51}

In the absence of photon counting noise, the observed ellipticity of a 
galaxy, $e^{\rm obs}_{\rm i}$, can be expressed as
\begin{equation} 
e^{\rm obs}_{\rm i}=e^{\rm s}_{\rm i}+P^{\gamma}_{\rm ij}\gamma_{\rm j}+P^{\rm sm}_{\rm ij}p_{\rm j}, 
\label{equation:ellips}
\end{equation} 
\noindent 
where $e^{\rm s}_{\rm i}$ is the intrinsic ellipticity of the galaxy, 
the second term is the shift in ellipticity caused by the gravitational shear, $\gamma_{\rm i}$,
and the third term describes the smearing of the galaxy image with an anisotropic 
PSF. 
The $P^{\gamma}$ and the $P^{\rm sm}$ factors are the shear and the smear 
polarizabilities, respectively, and are determined by the {\em getshapes} routine 
in {\sc imcat}, along with the ellipticities $e_{\rm i}^{\rm obs}$
(for details, see e.g.\ Kaiser et al.\ 1995; Clowe et al.\ 2000).

\begin{figure} 
\includegraphics{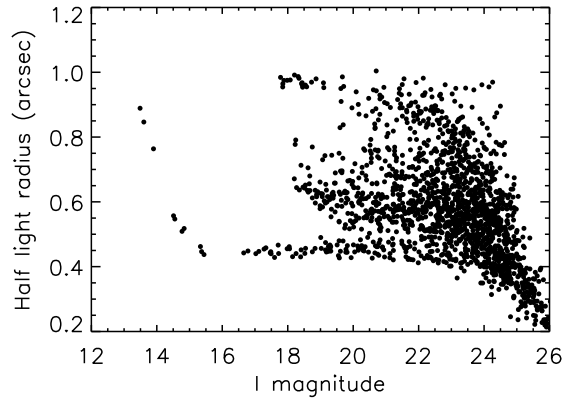} 
\caption{Half light radius versus magnitude of objects in the E1821$+$643 field. 
Stars are easily identified as the objects on the horizontal 
branch stretching from $I\approx15$ to 20--21. At $I \la 15$, the 
stars saturate and the branch turns upwards.}
\label{figure:fig4} 
\end{figure}  

By applying Eq.~\ref{equation:ellips} to stars which are intrinsically
circular ($e^{\rm s}_{\rm i}=0$) and not affected by 
gravitational shear ($\gamma_{\rm i}=0$), an estimate
of `stellar shear field', $p_{\rm i}$, can be made. 
To do this, we identified unsaturated, bright stars in the images 
by their half light radii as seen in Fig.~\ref{figure:fig4}. The stars 
were used to investigate the variation of the PSF across 
the field by fitting a second order polynomial to 
$p_{\rm i}$ using the {\sc imcat} task {\em efit}. 
The polynomial was thereafter applied to the galaxies by adjusting 
their ellipticities with an amount 
$\delta e_{\rm i} = -P^{\rm sm}_{\rm ij}p_{\rm j}$ using the task 
{\em ecorrect}. 

\begin{figure} 
\includegraphics{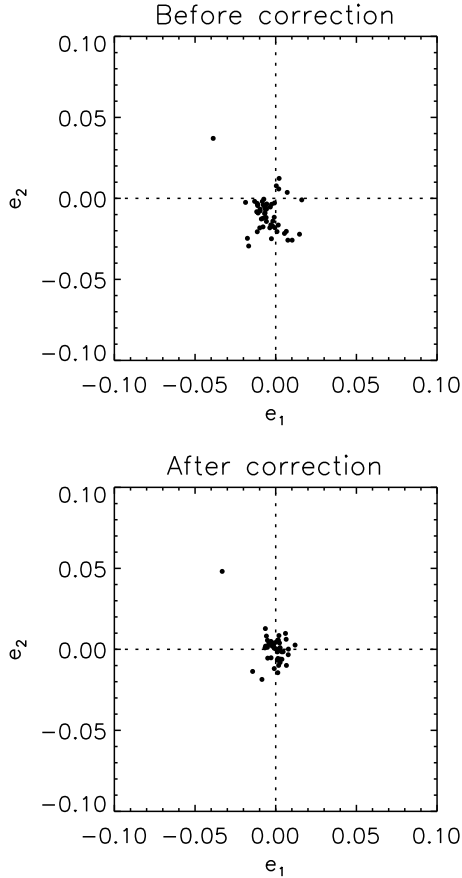} 
\caption{Ellipticities of stars before and after correction 
for PSF anisotropy
in the E1821$+$643 field. Perfectly circular stars have $e_{1}=e_{2}=0$.} 
\label{figure:fig5} 
\end{figure}

We experimented 
with higher order polynomials, but found that in most fields there were
too few stars to allow this. A second order polynomial
was however found to be satisfyingly correcting for any anisotropies. 
As an example, we show in Fig.~\ref{figure:fig5} the ellipticities 
of the stars in the 
E1821$+$643 field before and after the correction. 
This field has elliptical stars with systematic ellipticities
up to $\sim 2$--3 per cent in one direction, but it is seen that 
the systematics have been successfully corrected for, and that
the star ellipticities are reduced to typically $<1.5$ per cent.

In the 3C 295 field, we found systematic ellipticities of $<2$--3 per 
cent before correction.
The stars in the 3C 254 and the 3C 334 fields, which were taken 
on the same observing run, have systematic ellipticities 
of up to $\sim5$ per cent in one direction.
We had exceptionally good seeing during this
run, down to 0.4 arcsec, and experienced problems with focusing 
the telescope owing 
to an almost undersampled PSF. It is possible that the focussing problems 
might have caused the stars to become slightly elliptical.
There was also strong wind part of the time that might have caused
vibrations in the mirror and also given rise to the focussing problems.
However, in all the fields, we were able to successfully correct for 
the systematics with residual ellipticities of $<1$--2
per cent uniformly scattered around zero. 

In the next step, we corrected for 
the dilution of galaxy ellipticities caused by the smearing by 
the seeing disk. This step is done in order
to convert the observed ellipticities to a gravitational shear.
We followed the approach by Luppino \& Kaiser \shortcite{lk97}, which is 
also described elsewhere (e.g.\ Clowe et al.\ 2000).
The pre-seeing shear polarizability is defined as  
\begin{equation} 
P^{\gamma}_{\rm ij}= P^{\rm sh}_{\rm ij} - P^{\rm sm}_{\rm ik} \frac{P^{\rm sh\star}_{\rm kl}}{P^{\rm sm\star}_{\rm lj}}, 
\end{equation} 
\noindent 
where the asterisks denote values for stars.
Under the assumption that 
the PSF is close to circular after the correction, the off-diagonal 
elements of the polarizabilities are small compared to the 
diagonal elements, so the polarizabilities can be approximated by  
$P = \frac{1}{2}\left(P_{11}+P_{22}\right)$. We therefore calculated 
the average $P^{\rm sh\star}/P^{\rm sm\star}$ for stars as  
\begin{equation} 
\left<\frac{P^{\rm sh}}{P^{\rm sm}}\right>^{\star} = \frac{1}{N_{\rm stars}}\sum_{\rm stars}\frac{P^{\rm sh\star}_{11}+P^{\rm sh\star}_{22}}{P^{\rm sm\star}_{11}+P^{\rm sm\star}_{22}}, 
\end{equation} 
\noindent 
and evaluated the pre-seeing shear polarizability as 
\begin{equation} 
P^{\gamma} = \frac{1}{2}\left(P^{\rm sh}_{11}+P^{\rm sh}_{22}\right)-\frac{1}{2}\left(P^{\rm sm}_{11}+P^{\rm sm}_{22}\right)\left<\frac{P^{\rm sh}}{P^{\rm sm}}\right>^{\star}, 
\end{equation} 
\noindent 
where $P^{\rm sh}$ and $P^{\rm sm}$ are provided by 
{\sc imcat}. 

The gravitational shear was thereafter found by computing
\begin{equation}
\gamma_{\rm i} = \frac{e_{\rm i}}{P^{\gamma}},
\end{equation} 
\noindent
so in order to convert the galaxy 
ellipticities to a shear, a correction factor of $1/P^{\gamma}$ was
applied. Each galaxy has a $1/P^{\gamma}$ factor associated with
it, and in principle it is possible to correct for PSF dilution 
on a galaxy-by-galaxy basis. As this tends to be very noisy,
we followed the approach by Hoekstra et al. \shortcite{hoekstra98}
by evaluating the correction factors as a function of magnitude and 
galaxy size instead. We split the galaxies into bins in $r_g$ 
and magnitude such that approximately 20--40 galaxies fell in each bin,
and computed the median $1/P^{\gamma}$ for each bin.
Typically, the widths of the bins were 0.5--1.0 mag in magnitude 
and 0.2--0.5 pixels in 
size. 

The bins containing the faintest and smallest galaxies, those that 
are most affected by the seeing, had the largest correction factors, 
typically 8--12, whereas a 
value of unity was reached for larger galaxies which are 
less affected by seeing.
Since the galaxies with the largest correction factors
are also the ones with the poorest shape determinations, we 
assigned weights to the galaxies.
For each bin, a normalized weight inversely proportional to the dispersion in 
$P^{\gamma}$ was calculated, and during the whole analysis, galaxies 
with poorly determined $1/P^{\gamma}$ were given less weight. 

Due to the mass-sheet degeneracy (a uniform sheet of mass can be
added without altering the shear), the quantity that we measure from 
galaxy images is
in reality the reduced shear, $g$,
\begin{equation}
g=\frac{\gamma}{1-\kappa},
\end{equation} 
\noindent
i.e.\ a combination of shear, $\gamma$, 
and convergence, $\kappa$. Since we are working in the weak lensing limit where
$\kappa \ll 1$, we use the approximation that $g \approx \gamma$.
The shear field can therefore be directly inverted to give the 
dimensionless projected surface mass density of the cluster. 


\begin{figure*}
\includegraphics[scale=0.9]{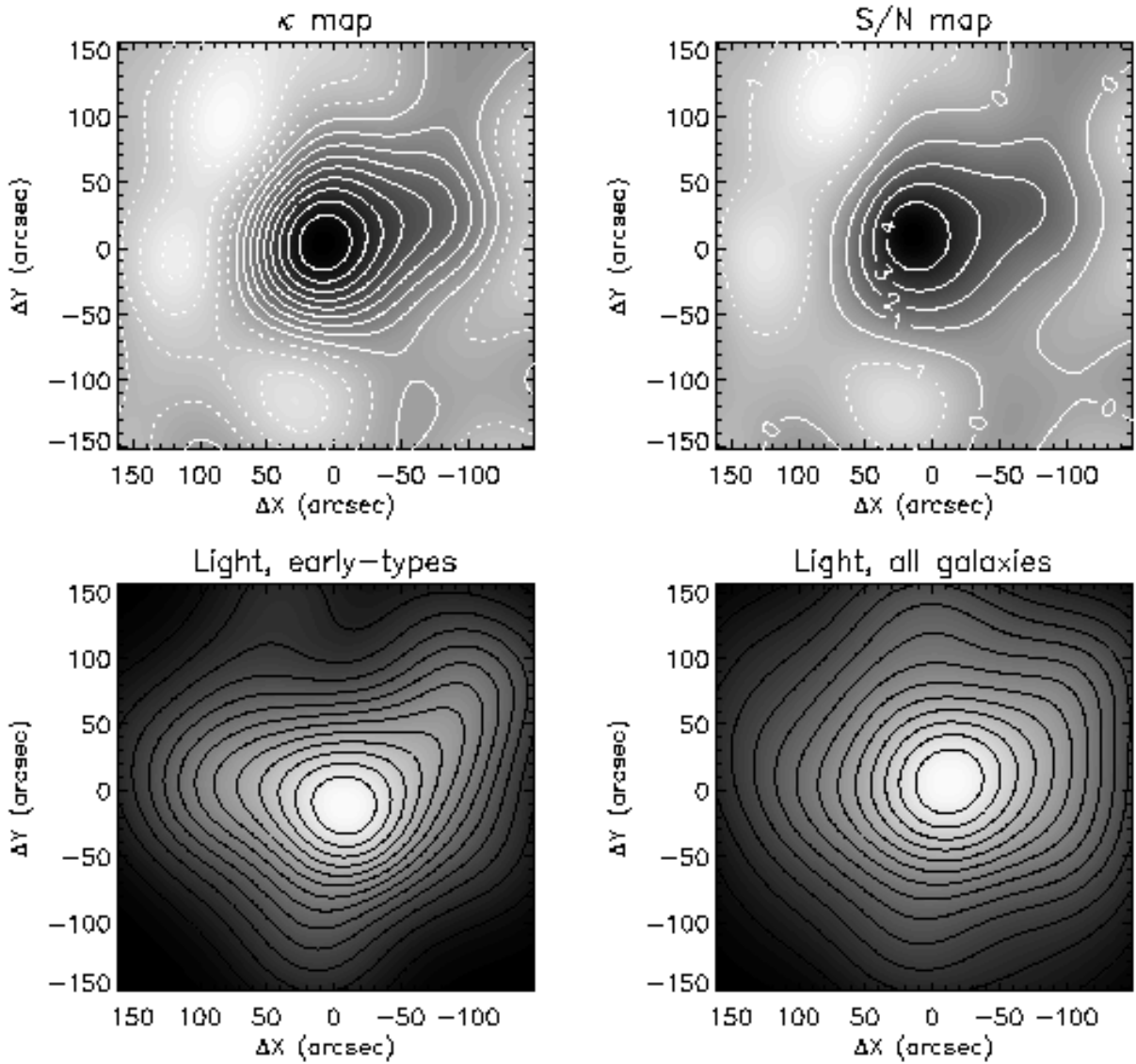}
\caption{E1821$+$643. The two upper panels show the results of the
mass reconstruction. To the left is the map of the projected surface
mass density, $\kappa$, and to the right, the signal-to-noise map of 
the reconstruction.
The two lower panels show the distribution of light from 
the colour-selected early-type galaxies (left) and from all the galaxies 
in the field (right).
In all panels, 60 grey levels were used, spanning from 
minimum to maximum. The spacing between the $\kappa$ contours
is 0.014, and the signal-to-noise contours are labeled with
their respective values. Negative contours are drawn with dotted lines.
The quasar position is $(\Delta x,\Delta y)=(0,0)$.
North is up and east is to the left.}
\label{figure:fig6}
\end{figure*}

\begin{figure*}
\includegraphics[scale=0.9]{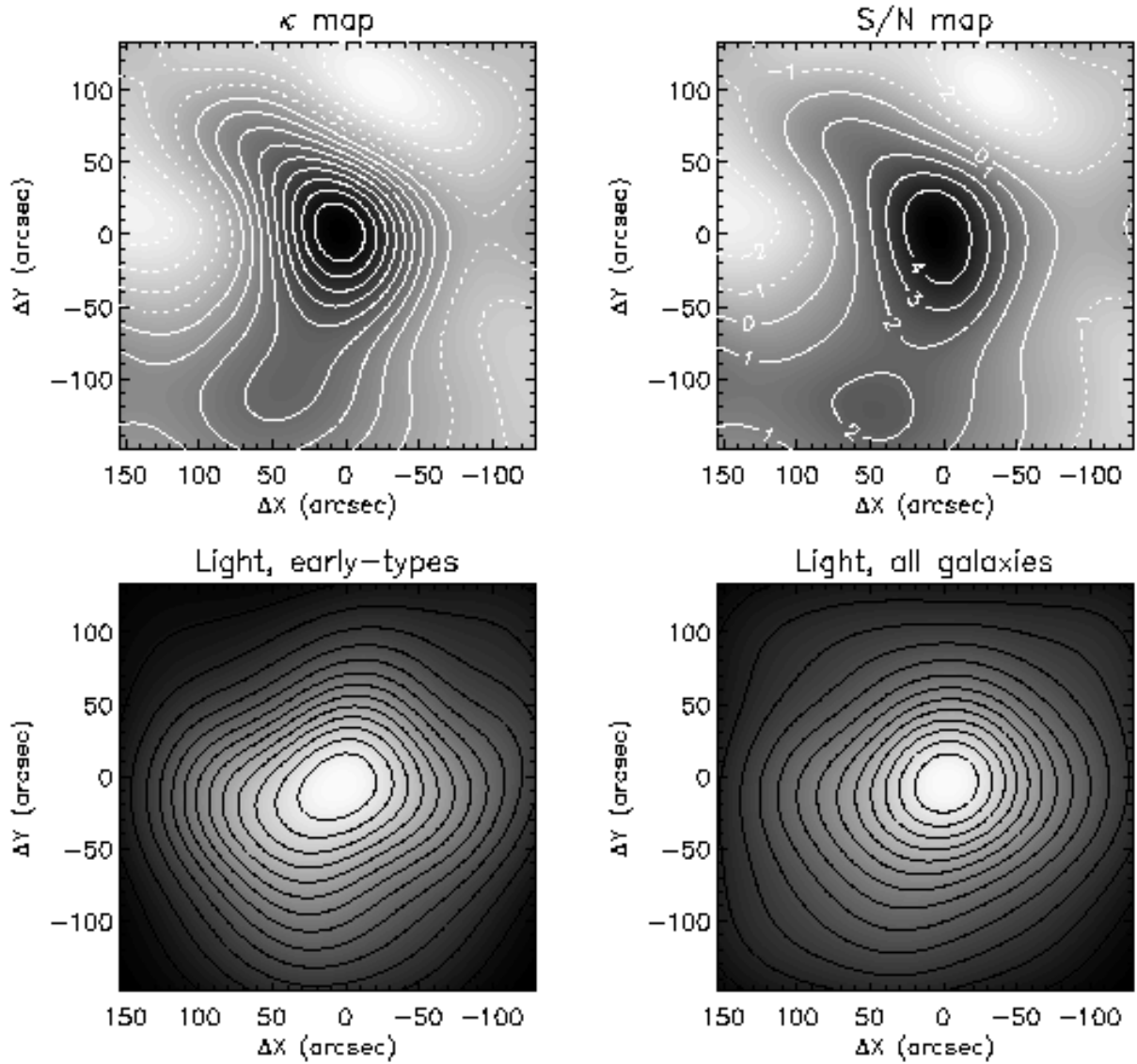}
\caption{3C 295. The radio galaxy has coordinates
$(\Delta x,\Delta y)=(0,0)$, and the spacing between the $\kappa$ contours is 0.013.
For more information about the plots, see text and caption of Fig.~\protect\ref{figure:fig6}.}
\label{figure:fig7}
\end{figure*}

\begin{figure*}
\includegraphics[scale=0.9]{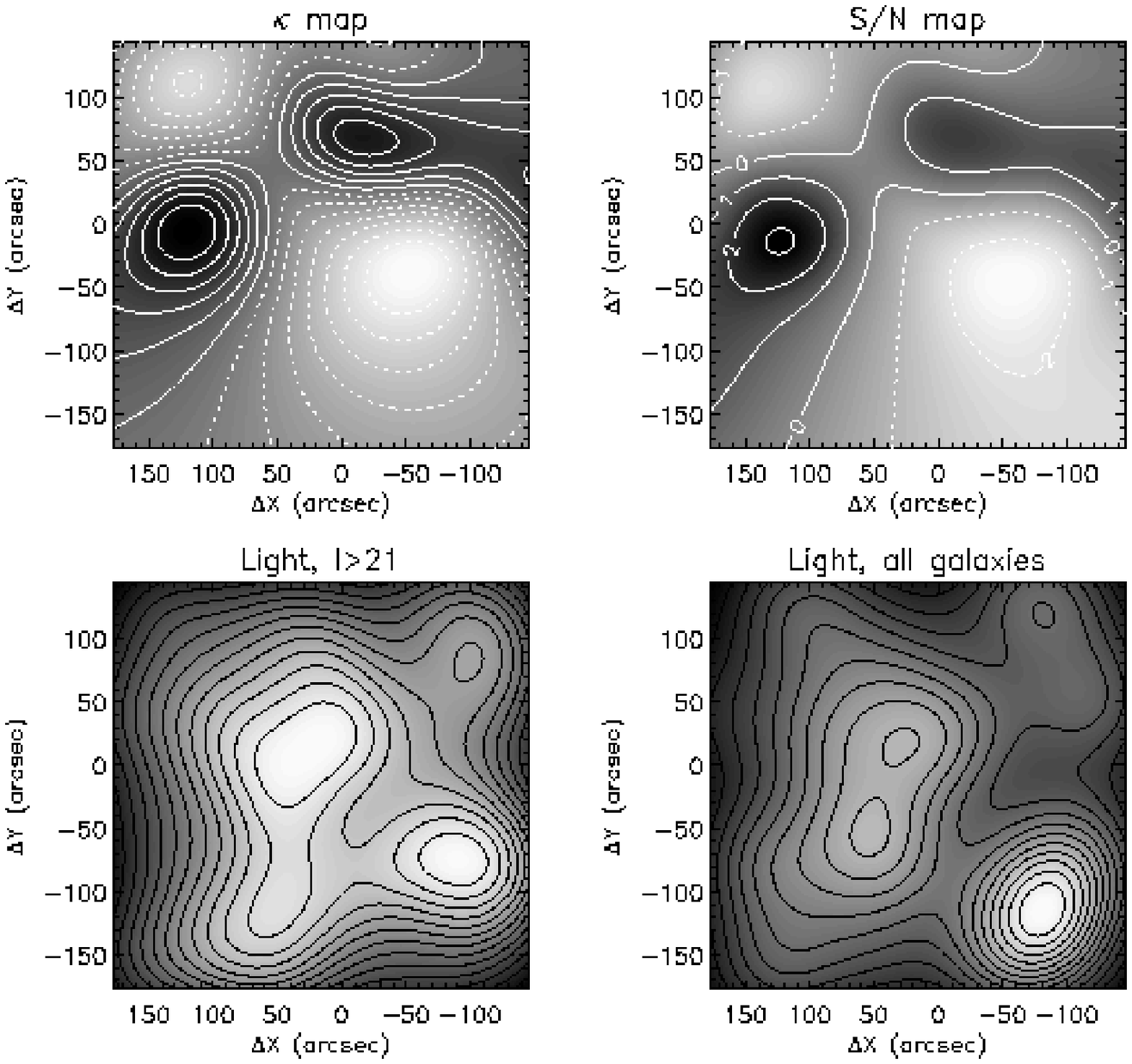}
\caption{3C 334. Surface mass density and light distribution, see caption 
of Fig.~\protect\ref{figure:fig6}. The quasar has coordinates
$(\Delta x,\Delta y)=(0,0)$, and the spacing between the $\kappa$ contours is 0.014.} 
\label{figure:fig8}
\end{figure*}

\subsection{Cluster inversion}
\label{section:s52}

We used the original `KS93' inversion technique 
\cite{ks93} to reconstruct
the surface mass density from the shear field.
The KS93 algorithm determines the $\kappa$ field to an unknown additive
constant and suffers somewhat from biasing at the edges
because of the finite field of view. However, it is
easy to implement and has well-defined 
noise properties since the intrinsic ellipticity distribution is translated
to white noise across the field. 
The surface mass density in each cluster was reconstructed on
an 82$\times$82 grid with a smoothing length of 200 pixels, corresponding 
to 38 arcsec. The typical resolution in the mass maps is therefore
$\approx 200-300 h^{-1}$ kpc at the cluster redshifts. 

To quantify the significance of the reconstructions
and to test the reliability of the mass maps, we made signal-to-noise 
maps of the mass reconstructions of each cluster.
This was done by generating 100 mass maps by randomly 
redistributing the measured shear values of the galaxies
while keeping their positions. A noise map was made by calculating
the {\it rms} in each reconstruction point, $\vec\theta$, as 
\begin{equation}
\kappa_{rms}\left(\vec\theta\right)=\sqrt{\frac{\sum_{\rm i=1}^{100} \kappa_{\rm i}\left(\vec\theta\right)^{2}}{100}}.
\end{equation}
\noindent
The noise maps show the expected noise level due to 
the random intrinsic ellipticities of the galaxies. Finally, 
a signal-to-noise map was made by dividing the original mass map with
the noise map. 

\begin{table}
\caption{The lens redshift, the weighted average $\left<\beta\right>$ and
the corresponding redshift, $z_{\rm sheet}$, for each field. The last
column lists the number density of galaxies used in the weak lensing
analysis.}
\begin{tabular}{lllll}

Field       & $z_{\rm lens}$ & $\left<\beta\right>$ & $z_{\rm sheet}$ & $n$ (arcmin$^{-2}$)\\
            &                &                    &   & \\
E1821$+$643 & 0.297 & 0.480 & 0.71 & 41.40\\
3C 295       & 0.460 & 0.322 & 0.80 & 50.65 \\
3C 334       & 0.555 & 0.175 & 0.73 & 19.86 \\
3C 254       & 0.734 & 0.236 & 1.13 & 41.57 \\

\end{tabular}
\label{table:tab3}
\end{table}


\section{Results}
\label{section:s6}

\subsection{Mass maps}

The mass reconstructions and their signal-to-noise maps
are shown in Figs.~\ref{figure:fig6}, \ref{figure:fig7}, \ref{figure:fig8} and \ref{figure:fig9},
in order of increasing quasar redshift. For comparison, 
we also show the distribution of light in the fields.
The early-type light distribution in the E1821$+$643 and the 3C 295 cluster
was calculated by selecting galaxies that lie within $1\sigma$ 
of the derived colour-magnitude relations,
whereas for the 3C 254 field, we selected galaxies in the colour range 
$2.5 < V-I < 3.5$. Next to the early-type light distributions, we
have plotted the light from all galaxies with $I>17$. 

\subsubsection{E1821$+$643}

As seen in Fig.~\ref{figure:fig6}, the cluster is comfortably
detected, and the mass peak has a signal-to-noise of 4.9.
The mass distribution peaks approximately 11 arcsec E of 
the quasar, but the offset is too small to be significant.
Probably, the peak is shifted because of random noise in 
the reconstruction.

The cluster appears to have a 
relatively smooth mass distribution. 
The mass contours show an extension to the 
NW which is present at a signal-to-noise level of 2--3 
(the resolution is $\approx200h^{-1}$ kpc). 
It is likely that this is a feature of the cluster, since the 
same asymmetry is present in the early-type light distribution. And, as 
expected, it is seen to be more
or less washed out when we plot the light distribution of all the galaxies
(the quasar host galaxy is not included in the light distribution
since it is saturated).
The {\em ROSAT} PSPC X-ray image published by Saxton et 
al.\ \shortcite{saxton97} appears to be 
slightly extended in the same direction as the mass map. An
apparent ellipticity of the cluster was also noticed in X-rays by 
Hall et al.\ \shortcite{heg97}, but it is not
clear whether this ellipticity corresponds to the feature 
seen in the mass map. A recent {\em Chandra} image
published by Fang et al.\ \shortcite{fang01} also shows 
ellipticity to some degree. 

We find a strong lensing candidate in this cluster. 
An elliptical galaxy which belongs
to the cluster (spectroscopically confirmed by Schneider 
et al.\ (1992), galaxy `H' in their paper)
has a close, very elongated neighbour 3 arcsec to the SE, see 
Fig.\ref{figure:e1821arc}.
The neighbour is slightly curved around the elliptical
galaxy and could thus be 
lensed by a combination of the cluster potential and the potential from the
cluster member.
By masking out the `lensed' galaxy and 
making a model of the elliptical cluster member using the 
{\em stsdas} tasks {\em ellipse} and {\em bmodel}, we  
subtracted the elliptical from the image and performed photometry
on the `lensed' galaxy. We found an $I$-magnitude of 21.09 and colours 
$V-I=3.15$ and $B-V=0.55$. Judging from the red $V-I$ colour, the 4000
{\AA} break probably lies at a redshifted wavelength of $\sim$8000 {\AA},
thus placing the galaxy at 
$z\sim1$. When properly modeled, cases like this 
can be used to put constraints on the mass distribution and dark matter
content in galaxies (e.g.\ Lubin et al.\ 2000).

\subsubsection{3C 295}

The distribution of the surface mass density and the galaxy 
light in the 3C 295 cluster is shown in Fig.~\ref{figure:fig7}. 
Also this cluster appears smooth 
in the mass map. 
Again, the mass peak is slightly offset ($\approx$7 arcsec
to the E) from the position of the central radio galaxy, but we attribute
this to random noise in the reconstruction. 
The peak is detected at a signal-to-noise level of 5.0, and the cluster
appears somewhat elongated in the NS direction.
Neumann \shortcite{neumann99}, who observed the cluster in X-rays with 
the {\em ROSAT} HRI also note the elongation, but 
in the {\em Chandra} X-ray image taken by Allen et al.\ \shortcite{allen01} 
this is not seen.

There is evidence for strong lensing also in this cluster.
We find a candidate gravitational arc 25 arcsec W 
and 10 arcsec N of the radio galaxy, the J2000 coordinates 
are 14:11:17.9 and $+$52:12:21.8.
It is detected in $I$, but not in the 
$V$ image. Smail et al.\ \shortcite{smail97a} make a note in their
paper that the cluster displays signs of strong lensing, but 
do not give the position of the arc. However, an arc is clearly
visible in the WFPC2 image at this position, so this must be the
arc found by Smail et al.
Gravitational arcs are usually found close to cluster centres at
approximately the Einstein radius.
If we take the Einstein radius to be at the position of the arc,
the Einstein radius is $\approx22$ arcsec, corresponding to 75$h^{-1}$ kpc. 
Assuming that the cluster core is a singular isothermal sphere and that
the arc lies at a redshift
of $z_{\rm source}=1.0$, the mass enclosed within the Einstein radius
is $1.02\times10^{14}h^{-1}$ M$_{\odot}$. For an arc redshift
of $z_{\rm source}=2$, the enclosed mass is $7.1\times10^{13}h^{-1}$ M$_{\odot}$.
Upon identification of any counter images and further modeling, 
tighter constraints can be made on the mass and the mass distribution 
in the cluster core. 

\subsubsection{3C 334}

The result of the cluster reconstruction in the 3C 334 field is 
shown in Fig.~\ref{figure:fig8} along with the light distribution of 
$I>21$ and $I>17$ galaxies.
As noted before, this field is not deep enough to obtain a
reliable weak lensing result. We applied the same selection criteria
in this field as for the other fields, except that we allowed
a slightly brighter magnitude cut at $I=20$ (we also tried with
$I=21$), but as seen in Table~\ref{table:tab3} there are very few galaxies
in this field compared to the other three fields. The completeness
limit is $I\approx23.5$, whereas a limit of $I\approx25$ is preferred
since most of the galaxies
contributing to the weak lensing signal have $I\approx23$--26.
Our current data can therefore neither
confirm nor exclude the presence of a cluster in this field.

The cloverleaf-like pattern in the mass map may originate in
the complex kernel in 
the reconstruction algorithm, and trying to reconstruct a 
mass map based on just one galaxy would produce a similar pattern.
However, the eastern peak in the mass map looks significant, so it
could be caused by a foreground mass concentration, but it is not supported
by the galaxy light distribution. It is possible that there is 
some systematic error in these data that we do not yet understand.  

\subsubsection{3C 254}

\begin{figure*}
\includegraphics[scale=0.85]{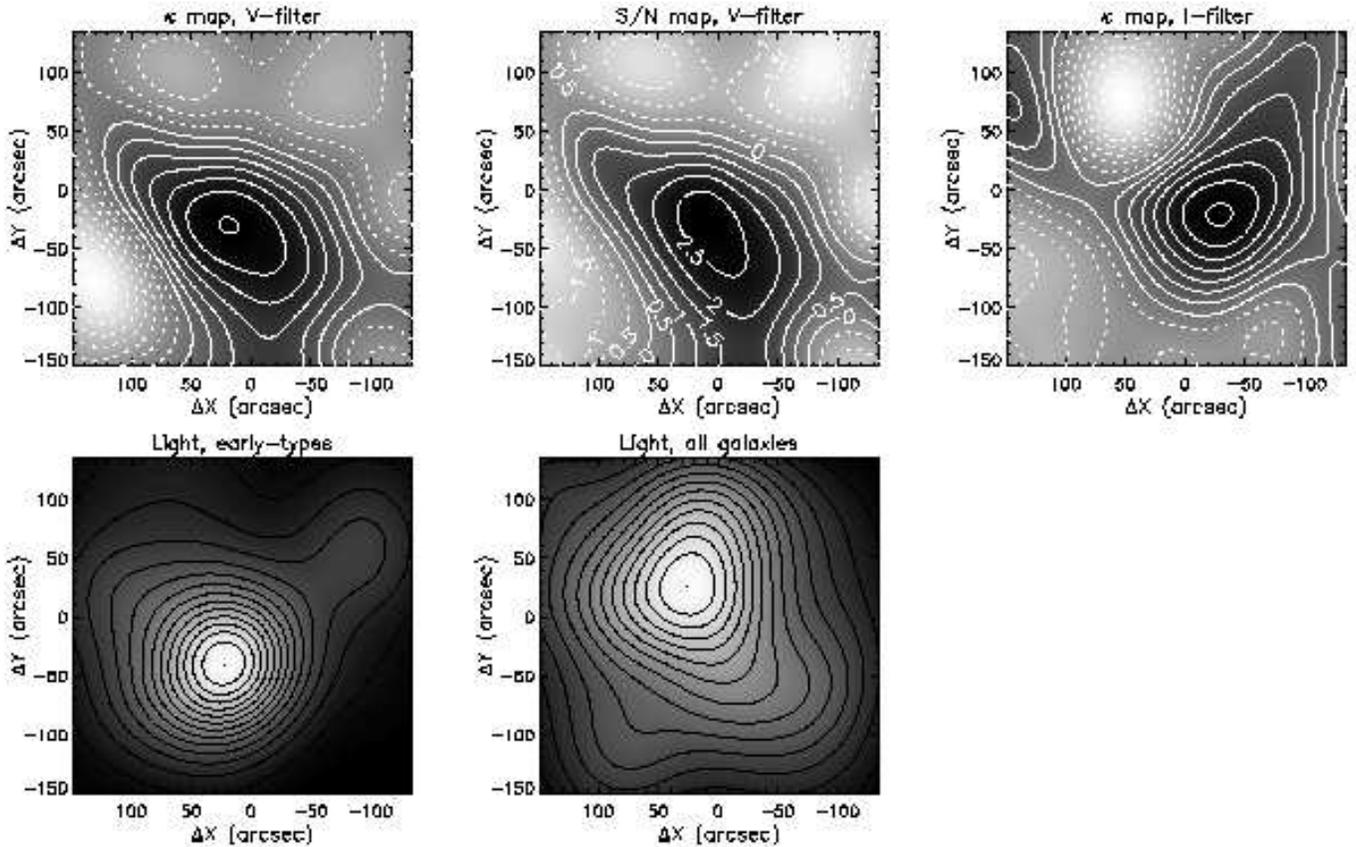}
\caption{3C 254. The upper row of panels shows, from left to right, the
$V$-band mass map, its signal-to-noise map, and the $I$-band 
mass map. The mass peak in $V$ has a 
signal-to-noise of 2.8, whereas the mass peak in $I$ has a signal-to-noise
of 1.9 (the $I$-band signal-to-noise map is not shown here). The spacing between the
$\kappa$ contours is 0.018 and 0.007 in the $V$- and the $I$-band
mass map, respectively.
In the two lower panels are shown the light distribution of 
colour-selected ($2.5 < V-I < 3.5$) galaxies and of all $I<21$ galaxies. 
The quasar has coordinates $(\Delta x,\Delta y)=(0,0)$.
See Fig.~\protect\ref{figure:fig6} for more details about the plots.}
\label{figure:fig9}
\end{figure*}

\begin{figure}
\includegraphics[scale=0.45]{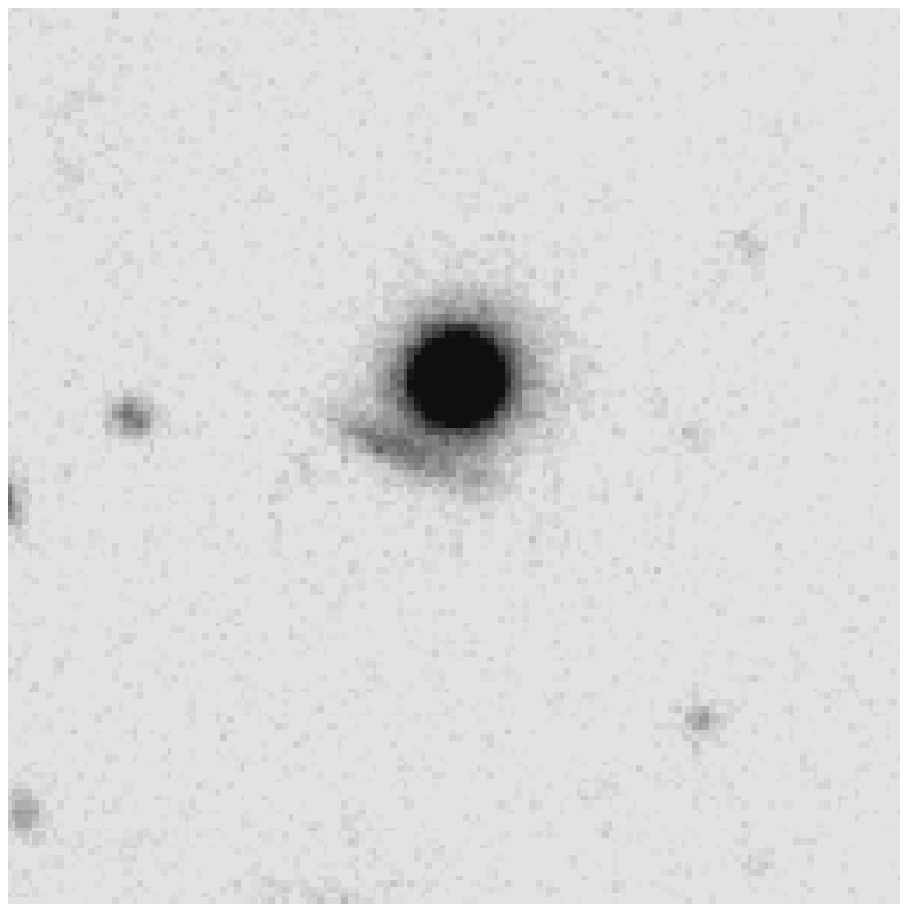}
\includegraphics[scale=0.45]{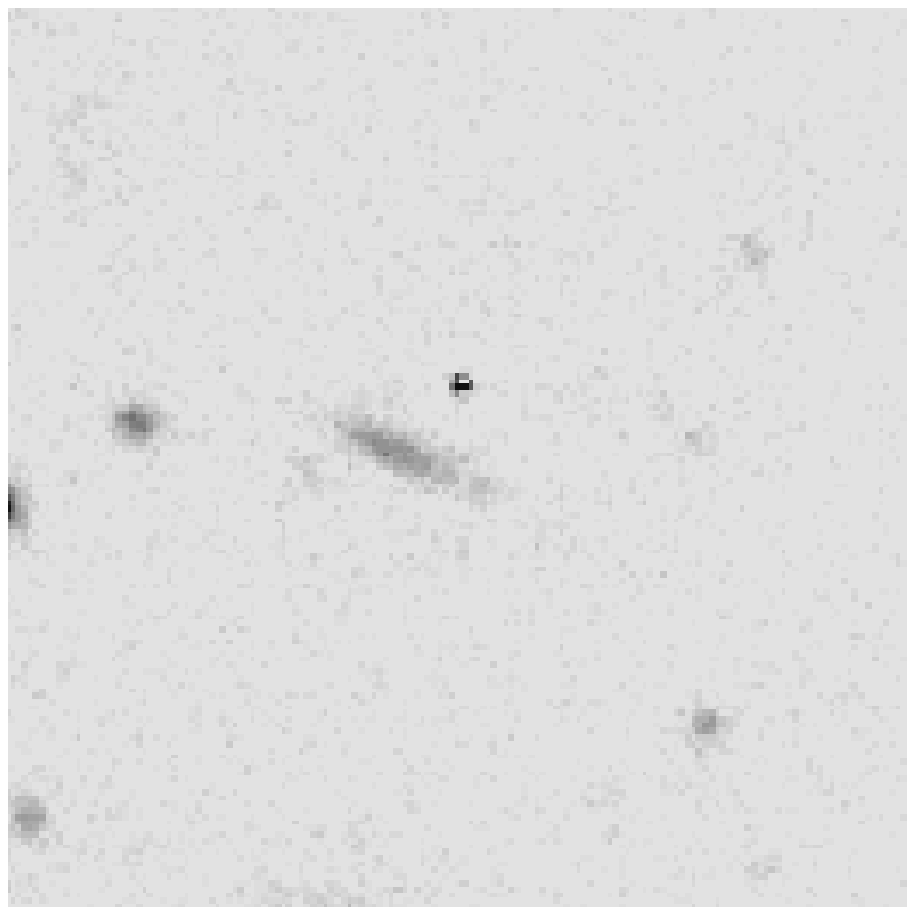}
\caption{A probable $z\sim1$ galaxy being lensed by a combination of 
the E1821$+$643 cluster potential and the gravitational potential from 
an elliptical cluster member. To the left is shown 
the elliptical galaxy with the candidate lensed galaxy $\approx$3 
arcsec to the SE (the J2000 coordinates of the `lensed' galaxy are 
18:21:59.6 and $+$64:19:47.6), and in the image to the right the
elliptical has been subtracted. 
North is up, and east is to the left.}
\label{figure:e1821arc}
\end{figure}

This is the most interesting field, both because of the relatively
high redshift of the quasar, and because there is a 
concentration of red galaxies in the image that may correspond to a cluster 
at the quasar redshift (see Section~\ref{section:s41}). 
As seen in Fig.~\ref{figure:fig9}, we have made a significant
detection of mass in this field. The mass peak in the $V$-band image
has a signal-to-noise of 2.8. Even though the $I$-band 
image is somewhat shallow with a
completeness limit of $\approx24$, there is also a detection of mass
in this image, although it is marginally significant with a signal-to-noise of
$\approx2$.

The $V$- and the $I$-band mass maps are seen to peak at different
places, approximately 47 arcsec apart, but we consider it unlikely 
that they originate in different mass concentrations. 
Taking into account the level of noise in the reconstructions,
they both probably correspond to the same cluster. 
The $V$-band mass map, which is the most reliable reconstruction, 
peaks 21 arcsec E and 34 arcsec S of the quasar. If the 
positional accuracy of the peak
is related to signal-to-noise as $\sim$smoothing length/signal-to-noise,
the 1-$\sigma$ error in the peak position is $\sim15$ arcsec. 
The position of the mass peak might therefore be consistent with 
the position of the quasar. 

We note that the mass peak lies closer 
to the peak in the early-type galaxy light; the separation between the two is 
$\approx10$ arcsec.
The early-type galaxy light 
corresponds to the concentration of red galaxies 
that was discussed in Section~\ref{section:s41} as a possible 
cluster at the quasar redshift. 
In this relatively noisy reconstruction it is unfortunately not 
possible to tell whether the cluster is centered on the quasar, or 
on the peak in the early-type light.

\subsection{Tangential shear and mass profiles}

We now derive the surface mass density and mass profiles of the clusters 
using the aperture mass densitometry method (Fahlman et al.\ 1994; Kaiser et al.\ 1994). 
This method converts the tangential component
of the shear to a surface mass density profile,
which is thereafter used to obtain a lower limit on the
mass as a function of radius.

\begin{figure*}
\includegraphics{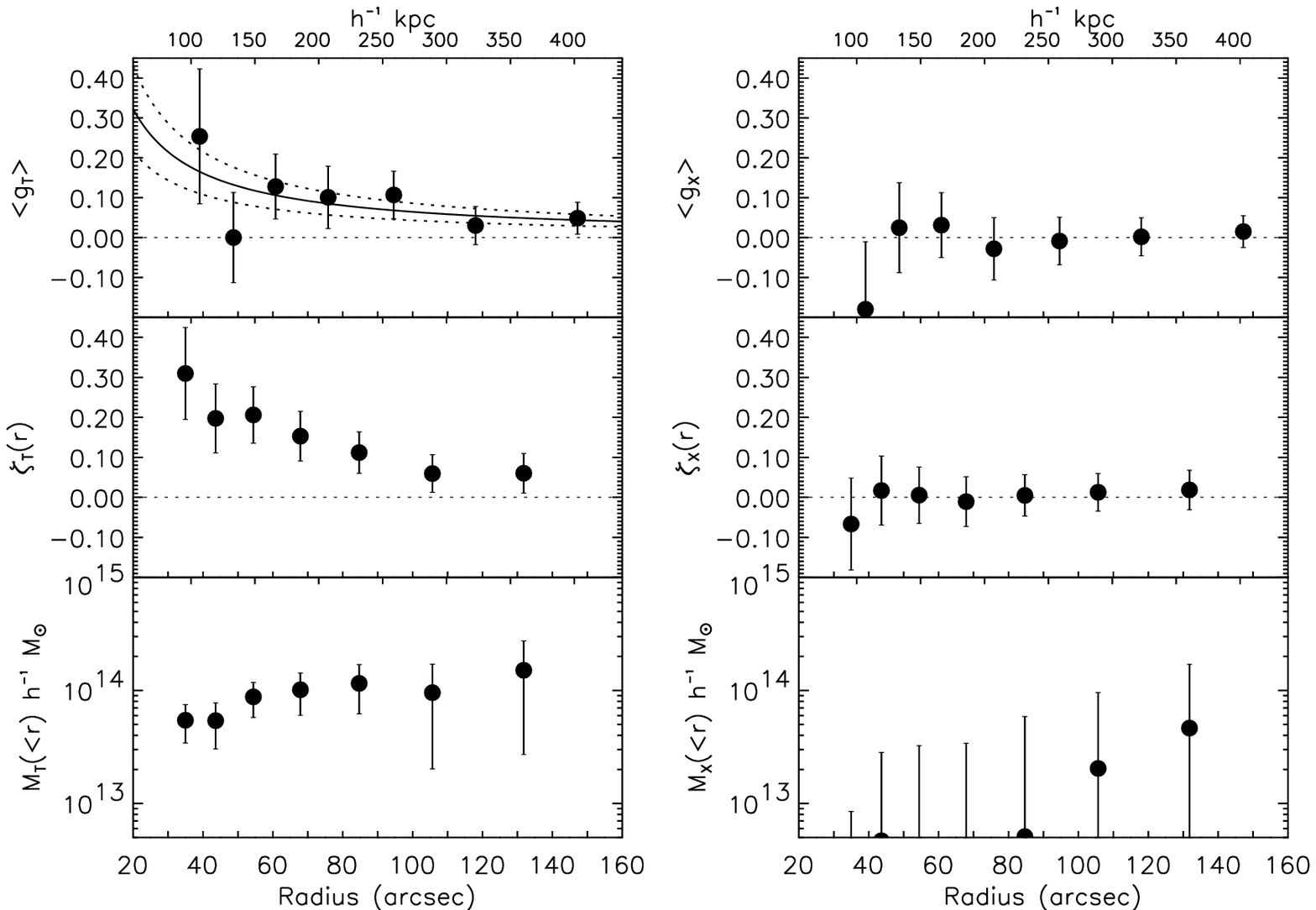}
\caption{E1821$+$643. The two upper plots show the tangential, 
$\left<g_{\rm T}\right>$, and the 
orthogonal shear, $\left<g_{\rm X}\right>$, as a function of radius. 
Below is shown their 
respective $\zeta$-estimators, denoted $\zeta_{\rm T}$ and $\zeta_{\rm X}$,
and the corresponding mass profiles.
The inner radius was taken to be 35 arcsec and the outer radius
of the control annulus was 164 arcsec.
The best-fit SIS model with velocity dispersion 964$^{+149}_{-176}$ 
km\,s$^{-1}$ is plotted as the solid and the two dashed 
lines in the upper left-hand plot.}
\label{figure:fig11}
\end{figure*}

First, we evaluated the azimuthally averaged tangential shear, 
$\left<g_{\rm T}\right>$, in annuli centered on the peak in the mass 
maps. The tangential
component of the shear is 
$g_{\rm T}=-g_{1}\cos(2\phi)-g_{2}\sin(2\phi)$, where
$\phi$ is the azimuthal angle of the galaxy with respect to the 
chosen centre. Thereafter, we calculated the $\zeta$-estimator 
which gives the mean dimensionless surface mass density within a 
radius $r_{\rm j}$ relative to that in an outer control annulus 
at $r_{\rm j} < r < r_{\rm max}$:
\begin{eqnarray}
\zeta(r_{\rm j},r_{\rm max}) & = & \bar\kappa(<r_{\rm j}) - \bar\kappa(r_{\rm j}<r<r_{\rm max}) \nonumber \\
                         & = & \frac{2}{(1-r_{\rm j}^{2}/r_{\rm max}^{2})} \int_{r_{\rm j}}^{r_{\rm max}} {\rm d}\ln r\left<g_{\rm T}\right>. 
\end{eqnarray}
The $\zeta$-statistic gives a lower limit to the mass within some radius
since the mass in the control annulus always will be different from zero.
In order to get a good measurement of the mass it is therefore desirable
to have a wide field so that $r_{\rm j}/r_{\rm max} \ll 1$. 

We used the peak in the mass maps as centre for the
annuli, and took the outer radius, $r_{\rm max}$, to be as large 
as possible without loosing 
too much of the area of the outer control annulus. The inner radius was
typically 35 arcsec, sufficiently outside the Einstein ring
for each cluster to assure that the weak lensing approximation 
was valid.
The $\zeta$-statistic is less well
applicable to fields with two or more peaks in the mass distribution
since the mass 
for one peak will be biased downwards when the other peak lies in 
the control annulus. 
The mass measured with this technique therefore depends both on the 
strength of the shear signal and on the mass profile of the cluster. 

\begin{figure*}
\includegraphics{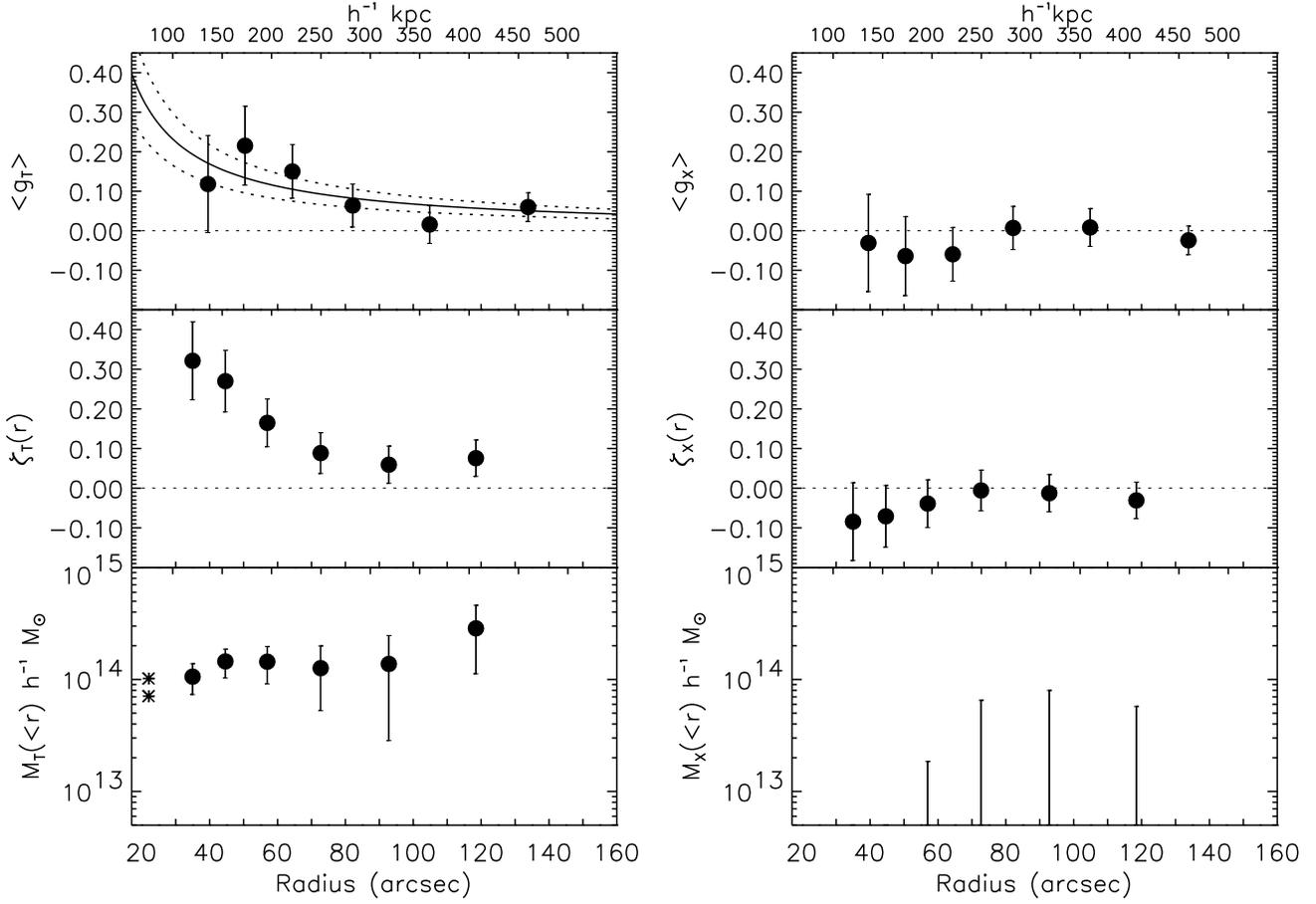}
\caption{3C 295. Tangential and orthogonal shear and their respective
$\zeta$-estimators and mass profiles. Inner and outer radii
were 35 and 151 arcsec, respectively. The two asterisks at a 
radius of 22 arcsec in the lower left-hand 
plot correspond to the mass estimated 
by strong lensing. The upper and lower asterisk symbols correspond to 
an assumed arc redshift of $z=1$ and $z=2$, respectively.
Overplotted on the shear profile, $\left<g_{\rm T}\right>$, is the best-fit 
SIS model with $\sigma_{v}=1205^{+161}_{-187}$
km\,s$^{-1}$ (solid and dashed lines).}
\label{figure:fig12}
\end{figure*}

\begin{figure*}
\includegraphics{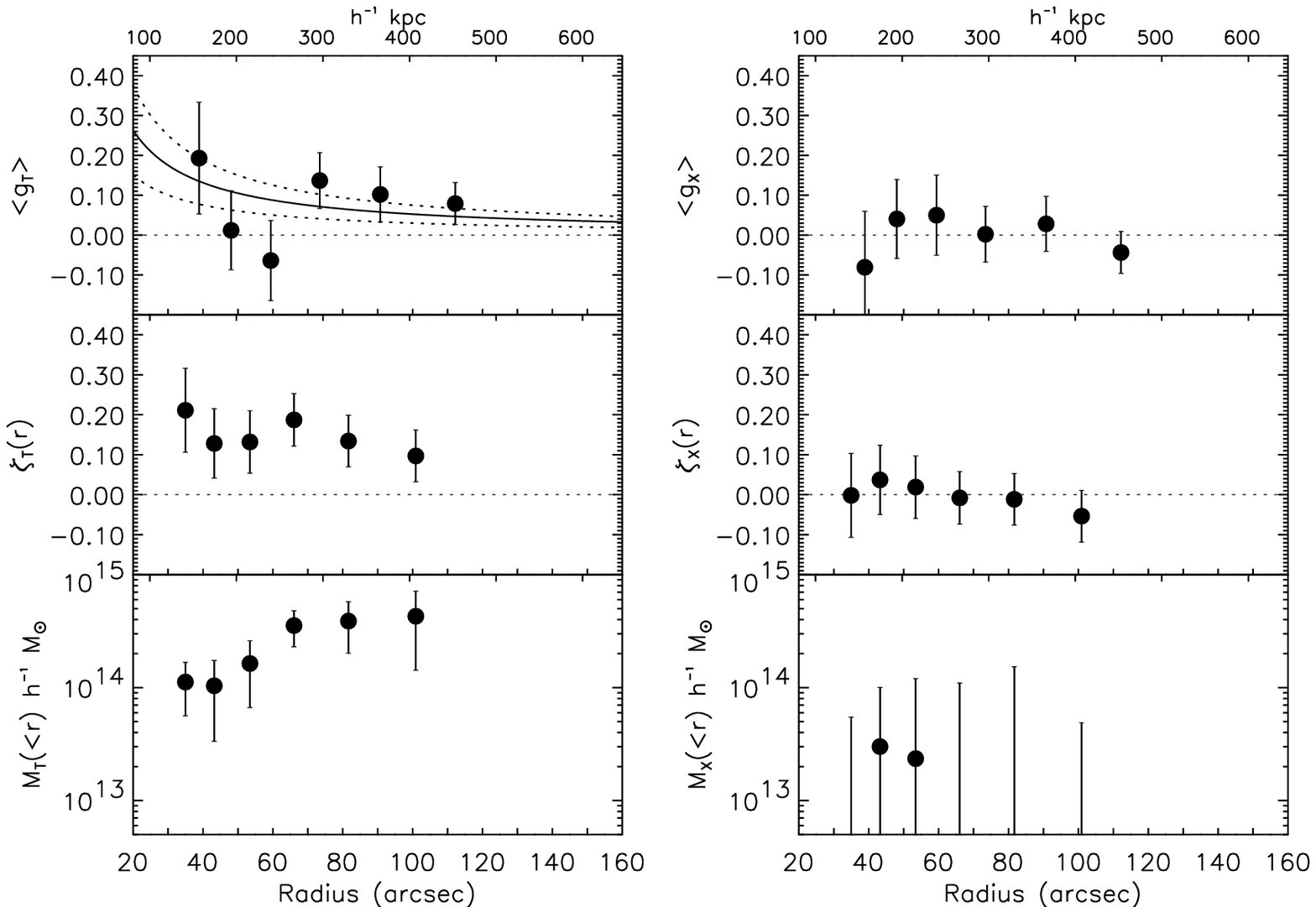}
\caption{3C 254. Tangential and orthogonal shear and their respective
$\zeta$-estimators and mass profiles.
The inner radius was 35 arcsec, and the outer radius
142 arcsec. The solid and the dashed lines in the upper left-hand
plot shows the best-fit SIS model with a velocity dispersion 
of $\sigma_{v}=1244^{+243}_{-302}$ km\,s$^{-1}$.} 
\label{figure:fig13}
\end{figure*}

In Figs.~\ref{figure:fig11} through \ref{figure:fig13}, we show
the results of the aperture mass densitometry and the mass profiles
for each cluster (from now on, we do not discuss the 3C 334 field).
The left-hand panels show, from top to bottom, the average tangential
shear profile, $\left<g_{\rm T}\right>$, its $\zeta$-estimator, $\zeta_{\rm T}$, and its
mass profile. The right-hand panels show the corresponding
quantities, but for the orthogonal shear component, $\left<g_{\rm X}\right>$.
The orthogonal shear was found by rotating the galaxies 
45 degrees, or equivalently,
increasing the phase of the shear by $\upi$ radians. 
It gives a measure of the
random shear component caused by the intrinsic ellipticities 
of the galaxies, and 
the variance in $g_{\rm X}$ was therefore used as a measure 
of the error in $g_{\rm T}$ \cite{lk97}. 
To ensure that the weak tangential shear originates in 
gravitational lensing and not some other systematic effect, a good
check is that the orthogonal shear component be zero within the statistical
uncertainty. Note that the 
points and the error bars for the $\zeta$-statistic (hence also the mass)
are correlated since $\zeta(r_{\rm j},r_{\rm max})$ gives the mean 
surface density interior to $r_{\rm j}$ 
relative to the mean in the outer control annulus. 
As seen in Figs.~\ref{figure:fig11} through \ref{figure:fig13},
there is a positive tangential shear detected in the three
clusters, and the orthogonal shear component
is scattered around zero, as expected if the weak shear is 
caused by gravitational lensing.

The mass profiles were calculated from the
$\zeta$-statistic by evaluating 
$M(<r_{\rm j})=\pi r_{\rm j}^{2}\zeta(r_{\rm j},r_{\rm max})\Sigma_{\rm c}$.
It is thus seen that the critical surface mass density, $\Sigma_{c}$, has to 
be known in order to convert to
an absolute measure of mass.
From Eq.~\ref{equation:eq1}, $\Sigma_{\rm c}$ can be seen to 
depend on the redshift distribution of the weakly lensed 
galaxies through the ratio $\beta \equiv D_{\rm ds} / D_{\rm s}$.
This ratio is in principle unknown, at least at magnitudes
fainter than $I\approx23$. We can however make a reasonable
estimate of the redshift distribution of faint galaxies
by using the photometric redshifts that are available in the 
Hubble Deep Field (HDF). 

For this, we used the HDF photometric redshift catalogue by 
Fern{\'a}ndez-Soto, Lanzetta \& Yahil (1999) which contains
$\approx1000$ galaxies, and followed the approach of
Dahle et al.\ \shortcite{dahle02} by comparing the 
number counts of galaxies in the cluster images with the number counts
in the HDF. 

The $F814W_{AB}$ magnitudes were first converted onto 
the STMAG system using the tabulated values in the HDF 
web pages\footnote{http://www.stsci.edu/ftp/science/hdf/logs/zeropoints.txt},
and thereafter to Cousins $I$ using the 
prescription in the WFPC2 Instrument Handbook \cite{biretta00}. 
Since we used the
$V$-magnitudes in the 3C 254 field, we also transformed 
from $F606W_{AB}$ to Johnson $V$ with the conversion 
$V-F606W_{AB}=-0.09$ (found by using the {\em stsdas.synphot} task to 
transform a galaxy with spectrum $F_{\nu}\propto \nu^{-1}$).

We grouped the galaxies in the cluster images and in the HDF into magnitude
bins of width 0.5 between the magnitude limit used 
in the weak lensing analysis and a faint limit
of $I=26$ or $V=27$. Galaxies fainter than this were counted in one bin. 
For each HDF magnitude bin, a mean $\beta$ was calculated using
the photometric redshifts. Thereafter, we weighted $\beta$ in each
bin with the normalized weights associated with 
the galaxies in the cluster images within the same bin.
The weighted $\beta$ values were finally added and a 
single weighted average 
$\left<\beta\right>$ was calculated for each cluster.
This is equivalent
to putting the source population at a sheet behind the cluster.
In Table~\ref{table:tab3}, we list the values of $\left<\beta\right>$
that we used, and the corresponding sheet-redshifts.

For the 3C 295 cluster, we find a mass of 
$(2.86 \pm 1.74) \times 10^{14}h^{-1}$ M$_{\odot}$ within a radius of
407$h^{-1}$ kpc. This agrees well with the mass estimate by Smail et al.\
\shortcite{smail97a} of 
$(2.35\pm0.38) \times 10^{14}$ $h^{-1}$ M$_{\odot}$  
within 400$h^{-1}$ kpc.
Using {\em Chandra} X-ray data, Allen et al.\ \shortcite{allen01} derive 
a mass of $1.05^{+0.40}_{-0.25} \times 10^{14}h^{-1}$ M$_{\odot}$ within the
same radius.
Our estimate is thus in better agreement with
Smail et al., but since our error bars are large, it is also consistent 
with the mass found by Allen et al.
Allen et al.\ discuss the discrepancy between
the X-ray and lensing mass as possibly originating
in mass along the line of sight
which is detected via weak lensing, but not in X-ray images.
There is however a possibility the outermost point 
in our mass profile of 3C 295 might be overestimated because of the
inclusion of the secondary peak to the south in the mass map. 
The mass within 300$h^{-1}$ kpc is
$(1.38 \pm 1.09) \times 10^{14}h^{-1}$ M$_{\odot}$, more in agreement with
Allen et al.'s X-ray mass estimate.

The E1821$+$643 cluster is massive too, with a mass of
$(1.51\pm1.24)\times 10^{14}h^{-1}$ M$_{\odot}$ within a radius of $\approx 400h^{-1}$
kpc, but the most massive cluster appears to be that in the 3C 254 field,
with a mass of $(4.28\pm2.86) \times 10^{14}h^{-1}$ M$_{\odot}$ within
$\approx 400h^{-1}$ kpc. The masses within radii of 200$h^{-1}$ and 400$h^{-1}$
kpc in the three clusters are listed in Table~\ref{table:tab4}. 

There are several factors which may give rise to systematic errors in
the mass. 
The value of $\left<\beta\right>$ affects the mass, both through the redshift
distribution of the faint background galaxies and through the assumed
cosmology. If the weakly lensed galaxies lie at systematically higher
redshifts than assumed here, $\left<\beta\right>$ will be larger, so 
the clusters should be less massive. For instance, if the source population
for the 3C 295 cluster is redshifted out to $z_{\rm sheet}=1$ instead 
of $z_{\rm sheet}=0.8$ as assumed, 
the mass within 400$h^{-1}$ kpc will
be $(2.25\pm1.36) \times 10^{14}h^{-1}$ M$_{\odot}$.
We have assumed an Einstein-deSitter
Universe with $q_{0}=0.5$ and $\Omega_{0}=1$, but in a flat Universe
with $\Omega_{\Lambda}=0.8$ and $\Omega_{\rm m}=0.2$, both $\left<\beta\right>$ 
and $D_{\rm d}$ will be larger by a factor of $\approx 1.2$--1.3, thus giving
masses smaller by a factor of $\approx1.2^{2}$--$1.3^{2}$. 

The tangential shear is given by 
$\left<g_{\rm T}\right> = -{\rm d}\kappa/{\rm d}\ln \theta$, and for
a singular isothermal sphere (SIS), the surface mass density, $\kappa$,
is proportional to the velocity dispersion, 
$\sigma_{v}^{2}$. The shear profiles
were therefore used to estimate the velocity dispersions of the clusters 
by minimizing $\chi^{2}$. SIS models with the best-fit velocity dispersions 
and their 68 per cent confidence intervals are overplotted on the tangential 
shear profiles in Figs.~\ref{figure:fig11} to \ref{figure:fig13}, and 
we also list the results of the fitting in Table~\ref{table:tab4}. 

The velocity dispersions we find are generally lower
than those measured spectroscopically. Dressler et al. \shortcite{dressler99}
measure $\sim 1630$ km\,s$^{-1}$ for the 3C 295 cluster, whereas we 
derive a best-fit SIS model with $\sigma_{v}=1205^{+161}_{-187}$ km\,s$^{-1}$.
Likewise, for the E1821$+$643 cluster, we find 
$\sigma_{v}=964^{+149}_{-176}$ km\,s$^{-1}$ 
from the SIS fit, but 1180 km\,s$^{-1}$ from spectroscopy of galaxies in the field
(Schneider et al.\ 1992; Le Brun et al.\ 1996; Tripp et al.\ 1998).
The discrepancy between the two methods of measuring velocity dispersions
was noted by Smail et al.\ \shortcite{smail97a}, who suggested 
it could be caused by different velocity dispersions for different 
galaxy populations. The early-type population has a lower velocity
dispersion and is more centrally concentrated
than the late-type population.
Spectroscopically measured velocity dispersions thus become overestimated
since they are often based on 
galaxies of different types lying at larger distances from the cluster core
than that probed by weak lensing. 
On the other hand, Irgens et al.\ (2002) found for a sample of 12 clusters with
redshifts between 0.15 and 0.33 that estimates of velocity dispersions
made from weak gravitational lensing and from direct spectroscopic
measurements are in good agreement, with only insignificant systematic
bias.
Using either technique, however, it is 
clear that the clusters studied here have very deep potential wells, and are at the 
upper end of the distribution of Abell cluster velocity dispersions
(e.g.\ Girardi et al.\ 1998).

\begin{table*}
\begin{minipage}{10cm}
\caption{Physical parameters derived 
for the AGN host clusters. In the second and third columns are
listed the mass within 200$h^{-1}$ and 400$h^{-1}$ kpc, and in 
column four and five, the result of the SIS fits to the
shear profiles.}

\begin{tabular}{lllll}

Cluster & \multicolumn{2}{c}{Mass (10$^{14}$ M$_{\odot}$)} & $\sigma_{v}$ (68 \% C.I.) & $\chi^{2}$/dof (dof) \\

  & $<200h^{-1}$ kpc & $<400h^{-1}$ kpc & km\,s$^{-1}$ & \\ 

                               & &     & & \\
E1821$+$643  & 1.01$\pm$0.41 & 1.51$\pm$1.24 & 964  (788,1113) & 0.41 (6) \\

3C 295        & 1.44$\pm$0.53 & 2.86$\pm$1.73 & 1205 (1018,1366) & 0.50 (5) \\

3C 254        & 1.63$\pm$0.97 & 4.28$\pm$2.86 & 1244 (942,1487) & 1.02 (5) \\

\end{tabular}
\label{table:tab4}
\end{minipage}
\end{table*}
 
\subsection{Mass-to-light ratios}

The M/L ratio parameterizes the amount of dark matter in
clusters. We have already measured the amount of total mass in 
the clusters, and by estimating the optical luminosity 
of the clusters, we may find their M/L ratios.
We processed the images in SExtractor with a tophat
convolution filter matched to the seeing, and for each extracted
object, aperture magnitudes within 2.6 arcsec and `best' magnitudes were 
evaluated. 
Stars were identified on the basis of a plot of FWHM versus 
the SExtractor star-galaxy classifier, and removed from the
catalogues by rejecting objects
that had a star-galaxy class greater than $\approx0.8$
and a FWHM corresponding to the seeing.
The best seeing catalogue was thereafter
matched to the catalogue in the other filter
on the basis of object positions, with
a tolerance of 3 pixels. A final catalogue was then made 
containing galaxies detected in both $V$ and $I$ 
(or $B$, $V$ and $I$ for E1821$+$643). 

The galaxy colours were calculated using the aperture
magnitudes, and the cluster early-type galaxies were again selected
by their location on the colour-magnitude relation. 
For consistency, we used the same 
sigma clipping algorithm as described in 
Section~\ref{section:s4} to define the colour-magnitude relations
from the SExtractor detections, but obtained very similar results
to those listed in Table~\ref{table:tab2}.

The $B$-band luminosity in solar units of each galaxy on 
the colour-magnitude sequence 
was evaluated using the `best' $I$ magnitudes converted to 
absolute $B$ magnitudes by first applying an $I$-band $K$-correction
for E/S0 galaxies
from Rocca-Volmerange \& Guiderdoni \shortcite{rvg88},
and thereafter adding the rest-frame $B-I=2.27$ of elliptical
galaxies (Fukugita, Shimasaku \& Ichikawa 1995). 
The $K$-corrections applied for the three clusters in order of
increasing redshift were 0.18, 0.39 and 0.635.
We converted to solar $B$ luminosity by adopting
$M_{B \odot}=5.48$.

The procedure to compute M/L ratios as a function of radius is 
similar to that described for the $\zeta$-statistic; the 
luminosity density within 
a given aperture is compared to the luminosity density within a control annulus. 

The argument for using the early-type galaxy light to calculate the
M/L ratio is that most of the light, and also most of the
mass, is associated with the early-type cluster galaxies. Kaiser et al.\ 
\shortcite{kaiser98} find that 70 per cent of
the total excess light within their cluster apertures comes
from early-type galaxies. They also argue that fairly accurate
estimates of cluster luminosities can be obtained from the early-type
cluster population since the noise due to foreground and background
contamination will be greatly reduced.
Smail et al.\ \shortcite{smail97a} also point out that using the 
early-type galaxy light for the M/L ratio 
makes it easier to compare with other clusters since the 
blue galaxy fraction is known to vary from cluster to cluster.

Nevertheless, we also calculated 
the M/L ratio obtained by selecting all galaxies at 
$I<17$.
By subtracting off the luminosity density in the control annulus,
any uniform component will be removed, thereby limiting the 
contamination by field galaxies. 
In doing this,
the galaxies on the colour-magnitude relation were converted to 
absolute $B$ luminosities as described above, whereas
the rest of the galaxies were $K$-corrected assuming an Sa-type 
galaxy at the cluster redshift and a rest frame colour of 
$B-I=1.99$ \cite{fukugita95}.
The Sa-type $K$-corrections we used were, in order of increasing
cluster redshift, 0.09, 0.22 and 0.38 \cite{rvg88}.

\begin{figure}
\includegraphics{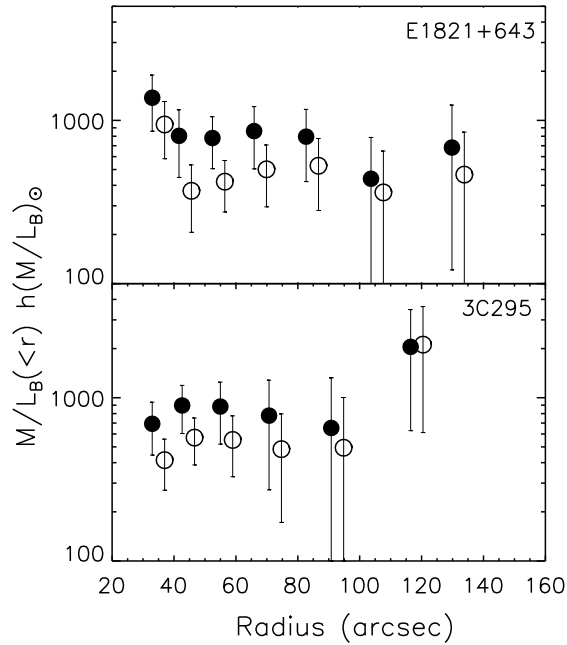}
\caption{M/L ratio as a function of radius.
Filled symbols correspond to early-type galaxy light and the open symbols
correspond to light from all galaxies having $I>17$. The points are
shifted slightly in the x-direction for clarity.}
\label{figure:fig14}
\end{figure}

The results for the 3C 295 and the E1821$+$643 clusters are shown graphically in 
Fig.~\ref{figure:fig14}. 
The errors in the M/L ratio were calculated assuming photometric errors of 
0.1 mag in addition to the 
previously given errors in the mass.
The early-type M/L ratios we find for the two clusters 3C 295 and
E1821$+$643 lie typically between $\approx$500$h$ and 
1000$h$ (M/L)$_{\odot}$ at the different radii, and the 
M/L ratios which include all galaxies are 
typically 300-500$h$ (M/L)$_{\odot}$, depending on the radius.
Smail et al.\ \shortcite{smail97a} get an early-type M/L ratio in $V$-band 
of 900$h$ (M/L)$_{\odot}$ within 400$h^{-1}$ kpc 
for the 3C 295 cluster. Assuming an average colour
of $B-V=0.78$ for the galaxies, corresponding to that of an
Sab-type galaxy (Fukugita et al.\ 1995), and adopting $(B-V)_{\odot}=0.68$,
we get that Smail et al.'s M/L ratio converts to 
$\approx987h$ (M/L)$_{\odot}$ in $B$-band. 
This agrees with our estimate of 2050$\pm$1420 (M/L)$_{\odot}$ 
within the same radius. 

In the 3C 254 field, we find 39 galaxies with $2.5 < V-I < 3.5$ 
within the radius of the outer control annulus. This is clearly
too few galaxies to obtain a reliable result with the 
method described above. The M/L ratio for the 3C 254 cluster is therefore
highly uncertain at this point. 

We note that the M/L ratios we find are moderately high. 
The mean early-type M/L ratio obtained by Smail et 
al. \shortcite{smail97a} for their ten clusters at
$z\approx0.2$--0.5 is (M/L$_{V}$)$^{\rm early}=483\pm103h$,
and the median is 410$h$ (M/L)$_{\odot}$. This corresponds to 
$\approx530h$ and $\approx450h$ in the $B$-band, so the values of 
$\approx700h$ we get are moderately high in comparison. 
When the whole galaxy population is included, it is found that
rich clusters where the early-type fraction is high have M/L ratios 
of typically $\sim 300$--400$h$ (M/L$_{B}$)$_{\odot}$ 
(Bahcall, Lubin \& Dorman 1995; Carlberg et al.\ 1997; 
Dahle et al.\ 2002, in prep.). Our corresponding number
for the E1821$+$643 and the 3C 295 cluster 
is (M/L$_{B}^{\rm all}$) $\sim 500h$ (M/L)$_{\odot}$.  

\section{Discussion}
\label{section:s7}

Our results for the 3C 295 cluster are in agreement with what
has been found for this cluster previously, i.e.\ the cluster 
is massive and has a high velocity dispersion, and the M/L ratio is
moderately high. 
It also shows signs of strong lensing
at a radius close to the Einstein radius we would predict from our
weak lensing measurements, suggesting that 
there is a large concentration of mass in the core 
($\sim10^{14}h^{-1}$ M$_{\odot}$) and
that the cluster potential is deep.

The mass distribution in the E1821$+$643 cluster is relatively smooth,
apart from some asymmetry to the NW. The same asymmetry is present in 
the early-type cluster light, indicating that the feature in the mass
map is real. The asymmetry could be 
caused by a group or a sub-cluster that has fallen in and merged with the
main cluster. If this is the case, the merger is probably in a fairly
advanced stage since the mass distribution is elongated rather
than double-peaked.
One of the cluster members in the direction of the asymmetry 
(41 arcsec N and 15 arcsec W of the quasar)
is an FRI radio galaxy. Since FRI galaxies are known be central
cluster galaxies (Longair \& Seldner 1979; Prestage \& Peacock 1988, 1989; 
Hill \& Lilly 1991), we speculate that a sub-cluster or a galaxy group
associated with the FRI source has merged with the main cluster.
Dynamical activity in clusters is relatively common, more than 30 per cent
of clusters show evidence in X-rays of merging \cite{jf92}. The
same fraction of activity is also found in weak lensing studies
\cite{dahle02}, suggesting that many clusters are still forming by
merging of sub-clusters and galaxy groups.

The morphology of the mass distribution in the 3C 254 field
is highly uncertain because the reconstruction is relatively noisy, but it appears to 
be a very massive cluster with a high velocity 
dispersion, $\sigma_{v} \approx 1200$ km\,s$^{-1}$.
Because of random noise, the position of the mass peak is 
also uncertain, 
but the position of the peak in the galaxy light distribution is 
easier to constrain. Galaxies with the expected colours
of ellipticals at the quasar redshift are concentrated
in a region 40 arcsec S of the quasar.
The $I$-band image reaches a completeness limit of $I\approx23.5$, corresponding to
$\approx2$ magnitudes fainter than $M^{*}$ at the quasar redshift, so the image
should be deep enough to trace the most massive
galaxies of the early-type population in a $z=0.734$ cluster. 

The 40 arcsec offset between the quasar and the
peak in the early-type galaxy light (assuming the colour-selected galaxies
lie at the quasar redshift) is not what we expect if the 
quasar sits in a massive cooling flow as suggested by 
Crawford \& Vanderriest (1997). A cooling flow cluster is expected to
be evolved and have a deep potential well, and the
early-type cluster galaxies should be centrally concentrated.
It would therefore be interesting to know where the mass map peaks in relation to the
quasar and the early-type galaxy light, but in our reconstruction, the positional 
errors are too large to address this.
  
Is it possible that foreground or background structure may
contaminate the morphology and the mass estimate of the cluster 
in the 3C 254 field? We found in 
Section~\ref{section:s41} what appears to be
a foreground group or cluster, possibly at $z\sim0.4$, but it lies
at a large angular distance from the quasar and is unlikely to have 
altered the morphology of the mass 
distribution around the quasar. Also, any background structure at a significantly 
larger redshift than the quasar would need to have an unrealistically high mass
to alter the reconstruction.

The evidence for an association of the detected mass
with the 3C 254 quasar is the following: 1) The mass distribution 
has a peak which is, within the uncertainties, centered on the quasar,
2) there is significant, extended X-ray emission around the 
quasar (Crawford et al.\ 1999; 
Hardcastle \& Worrall 1999), 3) the quasar is surrounded by an
extended emission-line region which signifies gas
under high pressure (Forbes et al.\ 1990; Crawford \& Vanderriest 1997), and
4) there is an apparent excess of galaxies around the quasar
\cite{bremer97}.
In the rest of the discussion we therefore assume that the cluster
is associated with the quasar, but not necessarily centered on it.

How do the clusters we have studied here fit into the interaction/merger
scenario and the cooling flow model for fuelling AGN in clusters?
Both the distribution of mass and galaxies in the clusters
detected at high signal-to-noise, E1821$+$643 and 3C 295, are 
fairly smooth, and their velocity dispersions are high. 
Unless there is sub-cluster merging along the line
of sight, this indicates that the clusters are evolved and dynamically 
relaxed. 
They therefore do not fit well into the scenario 
where galaxy-galaxy interactions
in dynamically young clusters fuel the AGN. 
Also, the moderately high M/L ratios we find seem to argue against 
this picture. One might expect that in a young non-virialized cluster, 
the cluster core would contain a smaller fraction of the total mass 
since the mass would still not have accumulated there, thus giving a
lower M/L ratio than a virialized cluster. 

Whereas the 3C 295 cluster is known to contain a cooling flow 
(Henry \& Henriksen 1986;
Neumann 1999; Allen et al.\ 2001), it is less certain that the
other two clusters have cooling flows.
So, although we cannot rule out 
the cooling flow model for the clusters, it seems 
possible that powerful AGN can exist in massive clusters with deep 
potential wells without being powered by cooling flows. 
E.g.\ Hall et al.\ \shortcite{heg97}
suggested that strong interactions between the AGN host and
a gas-rich galaxy is the only necessary mechanism for triggering
and fuelling AGN in clusters. 
However, as pointed out by Hall et al., this predicts that the 
AGN hosts have disturbed
morphology and that the clusters are dynamically young.

If instead interactions between galaxies can occur in dynamically 
relaxed clusters which have undergone mergers, this could  
explain our findings. If a merger between two galaxies can 
push gas into the centre of a galaxy, perhaps a merger between two clusters
could result in a whole galaxy being pushed towards the cluster centre.
A cluster merger event could thus disrupt the orbits of some galaxies and 
send them down into the centre of the potential well where the AGN host galaxy sits.
If the infalling galaxy is rich in gas, it could end up as a source of fuel
for the AGN.
Even if the dynamical friction is too low and the infalling
galaxy never interacts strongly with the AGN host, it could end up 
losing gas to the host, or at least disturb the 
gas in the host galaxy itself.
This scenario predicts that systems containing powerful AGN
show a high incident of cluster mergers. 

The M/L ratios of the clusters studied here seem to be moderately high 
compared to clusters in general. This could mean that they are indeed very 
massive and have high X-ray temperatures, typical for old clusters where 
the mass is more concentrated than the light \cite{bc02}. 
Another interesting possibility is that 
AGN-selected clusters have systematically higher M/L ratios than 
purely X-ray or optically selected clusters.
This might indicate that X-ray and optical surveys are biased toward clusters
with high baryon-to-dark matter ratios, and 
could be important to cosmology as it may affect estimates
of $\Omega$ in matter derived from analysis of known clusters
\cite{white93}. 


\section{Conclusions}
\label{section:s8}

Although our sample is too small to draw conclusions about the properties
of AGN host clusters in general, we can at least point to some interesting 
facts about clusters hosting powerful AGN:

\begin{enumerate}
\item They can be very massive systems with masses of a few times 10$^{14}h^{-1}$ M$_{\odot}$
within the central $\approx 500h^{-1}$ kpc. 
\item They can have very high velocity dispersions of 1000--1600 km\,s$^{-1}$
\item One shows evidence of a cluster merger event (E1821$+$643)
\item Even though the few clusters studied here are consistent with 
being drawn from the general cluster population, there is some tentative
evidence that they tend to have higher M/L ratios than clusters in general
\end{enumerate}
\noindent
The data presented here do not point unambiguously to either the cooling flow model or
the young cluster model for fuelling powerful AGN in clusters. 
Points i) and ii) above support the cooling flow model since cooling flows are
expected to occur in dynamically relaxed clusters with deep potential wells. 
On the other hand, point iii) argues against the cooling flow model, 
and more in 
favour of the young cluster model, although the
E1821$+$643 cluster is clearly relaxed. 

Our study has demonstrated that it is possible to find and 
study massive clusters associated with AGN using weak lensing techniques.
There seems to be good reasons for expecting moderately rich clusters
around the most luminous $z \la 1$ radio sources (e.g.\ Best 2000), and 
weak lensing observations in these fields might reveal whether they
are indeed surrounded by clusters. 


\section*{Acknowledgments}

The authors would like to thank the anonymous referee for a 
thorough reading of the manuscript, and for valuable comments
and suggestions which helped improve the presentation of this paper.
We are also grateful to the NOT staff for assistance during the
observations, and thank Magnus N{\"a}slund for discussions 
about luminosity functions. 
We also gratefully acknowledge Tomas Dahl{\'e}n for letting us use predictions
of galaxy colours from his photometric redshift code, and 
Brad Holden for estimating the velocity dispersion of the E1821$+$643
cluster.
MW acknowledges the Swedish Research Council for travel support. 

This research is based on observations made with the Nordic Optical
Telescope, which is operated on the island of LaPalma jointly by Denmark,
Finland, Iceland, Norway and Sweden, in the Spanish Observatorio del Roque 
de los Muchachos of the Instituto de Astrofisica de Canarias. The data presented here
have been taken using the ALFOSC, which is owned by the Instituto de Astrofisica
de Andalucia (IAA) and operated at the Nordic Optical Telescope under agreement
between IAA and NBIfAFG of the University of Copenhagen.

{\sc iraf} is distributed by the National Optical Astronomy Observatories, 
which are operated by the Association of Universities for Research in Astronomy,
Inc., under cooperative agreement with the National Science Foundation. 

This research has made use of the NASA/IPAC Extragalactic Database (NED), which is operated
by the Jet Propulsion Laboratory, California Institute of Technology, under contract
with the National Aeronautics and Space Administration. 


\end{document}